\documentclass[jcs]{llncs}
\usepackage{amsfonts}
\usepackage{amssymb}
\usepackage{graphicx}
\usepackage[dvipsnames]{xcolor}
\usepackage[caption=false,font=footnotesize]{subfig}
\usepackage{booktabs}
\usepackage{tabu}
\usepackage{paralist}
\usepackage{microtype}

\usepackage{hyperref}
\usepackage[all=normal,paragraphs=tight,bibbreaks=tight,floats=tight]{savetrees}

\newcommand{\atkname}{{\sf\em PILOT}}
\renewcommand{\paragraph}[1]{\medskip\noindent{\bf #1}}

\newcommand{\ignore}[1]{}

\newcommand{\new}[1]{#1}

\newcounter{inlinecounter}
\newcommand{\icstart}{\setcounter{inlinecounter}{0}}
\newcommand{\ic}{({\protect\refstepcounter{inlinecounter}\theinlinecounter)~}}

\begin{document}

\begin{frontmatter}
\title{\atkname: Password and PIN Information Leakage from Obfuscated Typing Videos \thanks{Submitted to the ESORICS 2018 special issue.}}

\author{Kiran Balagani\inst{1}\thanks{Authors are listed in alphabetical order.} \and
	Matteo Cardaioli\inst{2} \inst{4}\and
	Mauro Conti\inst{2} \and
	Paolo Gasti\inst{1} \and
	Martin Georgiev\inst{3}\thanks{Current affiliation: University of Oxford.} \and
	Tristan Gurtler\inst{1}\thanks{Current affiliation: University of Illinois at Urbana-Champaign.} \and
	Daniele Lain\inst{2}\thanks{Current affiliation: ETH Zurich.} \and
	Charissa Miller\inst{1}\thanks{Current affiliation: Rochester Institute of Technology.} \and
	Kendall Molas\inst{1} \and
	Nikita Samarin\inst{3}\thanks{Current affiliation: University of California, Berkeley.} \and
	Eugen Saraci\inst{2} \and
	Gene Tsudik\inst{3} \and
	Lynn Wu\inst{1}\thanks{Current affiliation: Bryn Mawr College.}
}



\institute{New York Institute of Technology \and University of Padua \and University of California, Irvine \and GFT Italy}

\maketitle

\begin{abstract}
	This paper studies leakage of user passwords and PINs based on
	observations of typing feedback on screens or from projectors in the form of masked characters ($*$ or $\bullet$)
	that indicate keystrokes. To this end, we developed an
	attack called \textit{Password and Pin Information Leakage from Obfuscated Typing Videos} (\atkname{}). Our attack extracts inter-keystroke timing
	information from videos of password masking characters displayed when
	users type their password on a computer, or their PIN at an ATM.
	We conducted several experiments in various attack scenarios.
	Results indicate that, while in some cases leakage is minor, it is quite
	substantial in others. By leveraging inter-keystroke timings, \atkname{}
	recovers 8-character alphanumeric passwords in as little as 19 attempts.
	When guessing PINs, \atkname{} significantly improved on both random guessing and the attack strategy adopted in our prior work~\cite{silktv}. In particular, we were able to guess about 3\% of the PINs within 10 attempts. This corresponds to a 26-fold improvement compared to random guessing. Our results strongly indicate that secure
	password masking GUIs must consider the information leakage identified
	in this paper.
\end{abstract}

\end{frontmatter}
\section{Introduction}
Passwords and PINs are susceptible to shoulder surfing
attacks~\cite{tari2006comparison} of which there are two main types:~(1)~input-based and (2) output-based. The former is more common; in it, the
adversary observes an input device (keyboard or keypad) as the user enters a
secret (password or PIN) and learns the key-presses. The latter involves the
adversary observing an output device (screen or projector) while the user enters
a secret which is displayed in cleartext. The principal distinction between the
two types is adversary's proximity: observing input devices requires the
adversary to be closer to the victim than observing output devices, which tend
to have larger form factors, i.e., physical dimensions.

Completely disabling on-screen feedback during password/PIN entry (as in, e.g., Unix
{\tt sudo} command) mitigates output-based shoulder-surfing attacks. Unfortunately, it also
impacts usability: when deprived of visual feedback, users cannot determine
whether a given key-press was registered and are thus more apt to make mistakes.
In order to balance security and usability, user interfaces typically implement
password masking by displaying a generic symbol (e.g., ``$\bullet$'' or ``$*$'')
after each keystroke. This technique is commonly used on desktops, laptops and
smartphones as well as on public devices, such as Automated Teller Machines
(ATMs) or Point-of-Sale (PoS) terminals at shops or gas stations.

Despite the popularity of password masking, little has been done to quantify
how visual keystroke feedback impacts security. In particular, masking assumes
that showing generic symbols does not reveal any information about the
corresponding secret. This assumption seems reasonable, since visual
representation of a generic symbol is independent of the key-press. However, in
this paper we show that this assumption is incorrect. By leveraging precise
inter-keystroke timing information leaked by the appearance of each masking
symbol, we show that the adversary can significantly narrow down the password/PIN's search space. Put another way, the number of attempts required to
brute-force decreases appreciably when the adversary has access to
inter-keystroke timing information.

There are many realistic settings where visual inter-keystroke timing
information (leaked via appearance of masking symbols) is readily available while
the input information is not, i.e., the input device is not easily observable.
For example, in a typical lecture or classroom scenario, the presenter's
keyboard is usually out of sight, while the external projector display is
wide-open for recording. Similarly, in a multi-person office scenario, an
adversarial co-worker can surreptitiously record the victim's screen. The same
holds in public scenarios, such as PoS terminals and ATMs, where displays
(though smaller) tend to be easier to observe and record than entry keypads.

In this paper we consider two representative scenarios: \icstart \ic a presenter
enters a password into a computer connected to an external projector; \ic a user
enters a PIN at an ATM in a public location. The adversary is assumed to record
keystroke feedback from the projector display or an ATM screen using a dedicated
video camera or a smartphone. We note that a human adversary does not need to
be present during the attack: recording might be done via an existing camera
either pre-installed or pre-compromised by the adversary, possibly remotely,
e.g., as in the infamous Mirai botnet~\cite{kolias2017ddos}.

\paragraph{Contributions.}
The main goal of this paper is to quantify the amount of information leaked
through video recordings of on-screen keystroke feedback. To this end, we
conducted extensive data collection experiments that involved 84
subjects.\footnote{Where required, IRB approvals were duly obtained prior to the
experiments.} Each subject was asked to type passwords or PINs while the screen
or projector was video-recorded using either a commodity video camera and a
smartphone camera. Based on this, we determined the key statistical properties
of resulting data, and set up an attack, called \atkname: \underline{P}assword and Pin \underline{I}nformation \underline{L}eakage from \underline{O}bfuscated \underline{T}yping Videos. It allows us to quantify reduction in
brute-force search space due to timing information. \atkname{} leverages
multiple publicly available typing datasets to extract population timings, and
applies this information to inter-keystroke timings extracted from videos.

Our results show that video recordings can be effective in extracting precise
inter-keystroke timing information. Experiments show that \atkname{}
substantially reduces the search space for each password, even when the
adversary has no access to user-specific keystroke templates. When run on
passwords, \atkname{} performed better than random guessing between 87\% and
100\% of the time, depending on the password and the machine learning technique
used to instantiate the attack. The resulting average speedup is between 25\% and 385\% (depending on the password), compared to random dictionary-based
guessing; some passwords were correctly guessed in as few as 68 attempts. A
single password timing disclosure is enough for \atkname{} to successfully
achieve these results. However, when the adversary observes the user entering
the password three times, \atkname{} can crack the password in as few as 19
attempts. Clearly, \atkname~'s benefits depend in part on the strength of a
specific password. With very common passwords, benefits of \atkname{} are
limited. Meanwhile, we show that \atkname{} substantially outperforms random
guessing with less common passwords. With PINs, disclosure of
timing poses an effective risk. The PIN guessing algorithm can reduce the number of
attempts up to 26 times compared to random guessing.

\paragraph{Paper Organization.} Section~\ref{sec:related_work} overviews
state-of-the-art in password guessing based on timing attacks.
Section~\ref{sec:systemthreat} presents \atkname{} and the adversary model. Section~\ref{sec:atkname} discusses our data collection and experiments. We then
present the results on password guessing using \atkname{} in
Section~\ref{sec:attack_passwords}, and on PIN guessing in
Section~\ref{sec:attack_pins}. The paper concludes with the summary and future work
directions in Section~\ref{sec:conclusion}.

\section{Related Work}\label{sec:related_work}
There is a large body of prior work on timing attacks in the context of
keyboard-based password entry. Song et al.~\cite{song2001timing} demonstrated a
weakness that allows the adversary to extract information about passwords typed
during SSH sessions. The attack relies on the fact that, to minimize latency,
SSH transmits each keystroke immediately after entry, in a separate IP packet.
By eavesdropping on such packets, the adversary can collect accurate
inter-keystroke timing information. Authors in~\cite{song2001timing} showed that this
information can be used to restrict the search space of passwords. The impact of
this work is significant, because it shows the power of timing attacks on
cracking passwords.

There are several studies of keystroke inference from analysis of video
recordings. Balzarotti et al.~\cite{balzarotti2008clearshot} addressed the
typical shoulder-surfing scenario, where a camera tracks hand and finger
movements on the keyboard. Text was automatically reconstructed from resulting
videos. Similarly, Xu et al.~\cite{xu2013seeing} recorded user's finger
movements on mobile devices to infer keystroke information. Unfortunately,
neither attack applies to our sample scenarios, where the keyboard is invisible
to the adversary.

Shukla et al.~\cite{shukla2014beware} showed that text can be inferred even from
videos where the keyboard/keypad is not visible. This attack involved analyzing
video recordings of the back of the user's hand holding a smartphone in order to
infer which location on the screen is tapped. By observing the motion of the
user's hand, the path of the finger across the screen can be reconstructed,
which yields the typed text. In a similar attack, Sun et
al.~\cite{sun2016visible} successfully reconstructed text typed on tablets by
recording and analyzing the tablet's movements, rather than movements of the
user's hands. 

The closest work to the paper is our prior work \cite{silktv}, in which we show that passwords can be inferred at a higher probability than random guesses using the timing information from onscreen keystroke feedback. However, in \cite{silktv} we concluded that the timing information is not helpful in inferring PINs. In this paper, we revisit our earlier conclusion on inferring PINs and show that it is incorrect. In fact, the attack strategy employed in this paper yielded a 26-fold improvement in inferring PINs over random guesses and significantly outperform [4] in terms of number of PINs recovered within a small number of attempts.  

Another line of work aimed to quantify keystroke information inadvertently
leaked by motion sensors. Owusu et al.~\cite{owusu2012accessory} studied this in
the context of a smartphone's inertial sensors while the user types using the
on-screen keyboard. The application used to implement this attack does not
require special privileges, since modern smartphone operating systems do not
require explicit authorization to access inertial sensors data.
Similarly, Wang et al.~\cite{wang2016friend} explored keystroke information
leakage from inertial sensors on wearable devices, e.g., smartwatches and
fitness trackers. By estimating the motion of a wearable device placed on the
wrist of the user, movements of the user's hand over a keyboard can be inferred.
This allows learning which keys were pressed during the hand's path.
Compared to our work, both \cite{owusu2012accessory} and \cite{wang2016friend}
require a substantially higher level of access to the user's device. To collect
data from inertial sensors the adversary must have previously succeeded in
deceiving the user into installing a malicious application, or otherwise
compromised the user's device. In contrast, \atkname{} is a fully passive
attack.

Acoustic emanations represent another effective side-channel for keystroke
inference. This class of attacks is based on the observation that different
keyboard keys emit subtly different sounds when pressed. This information can be
captured (1) locally, using microphones placed near the
keyboard~\cite{asonov2004keyboard,zhuang2009keyboard}, or (2) remotely, via
Voice-over-IP~\cite{compagno2017don}. Also, acoustic emanations captured using
multiple microphones can be used to extract locations of keys on a keyboard. As
shown by Zhou et al.~\cite{Zhu2014}, recordings from multiple microphones can be
used to accurately quantify {\em time difference of arrival (TDoA)}, and thus
triangulate positions of pressed keys.

\section{System and Adversary Model}\label{sec:systemthreat}
We now present the system and adversary model used in the rest of the paper. 

\begin{figure}[ht]
	\centering
	\begin{minipage}{.45\textwidth}
		\centering
		\includegraphics[width=.7\linewidth]{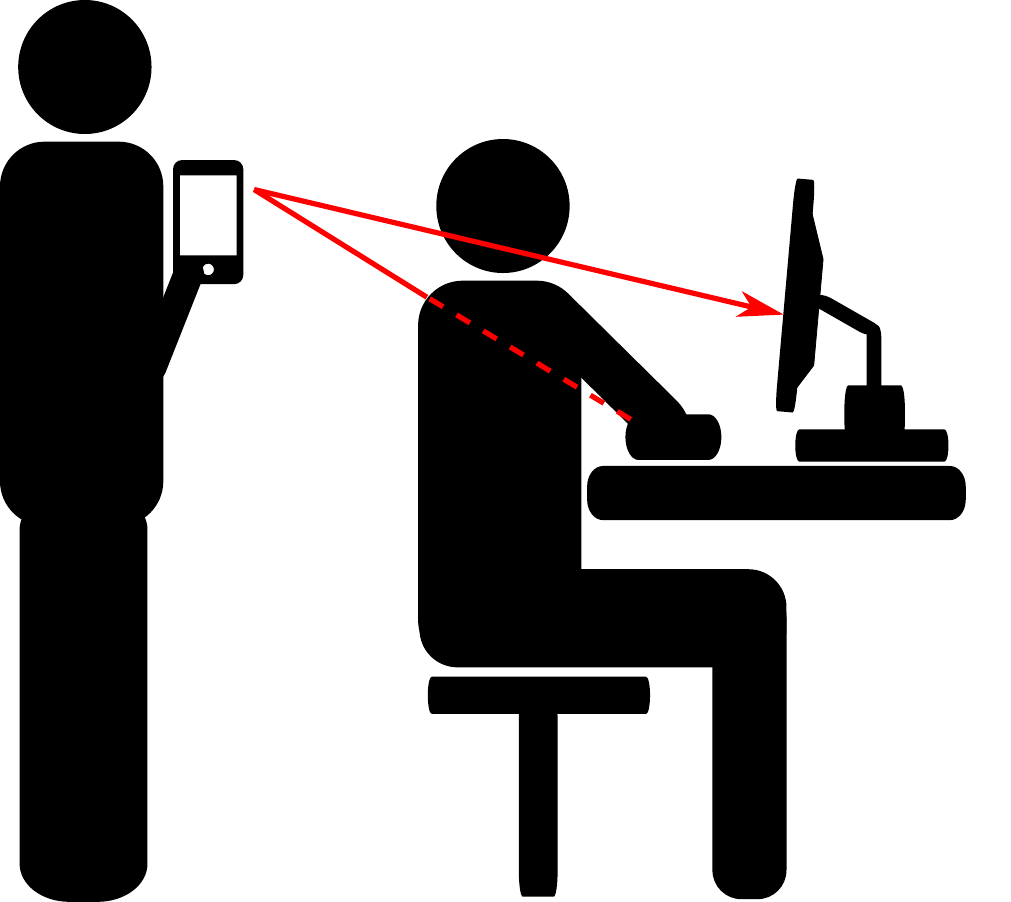}
		\caption{Example attack scenario.}
		\label{fig:shouldersurfing}
	\end{minipage}
	\begin{minipage}{.45\textwidth}
		\centering
		\subfloat[]{\includegraphics[width=0.48\linewidth]{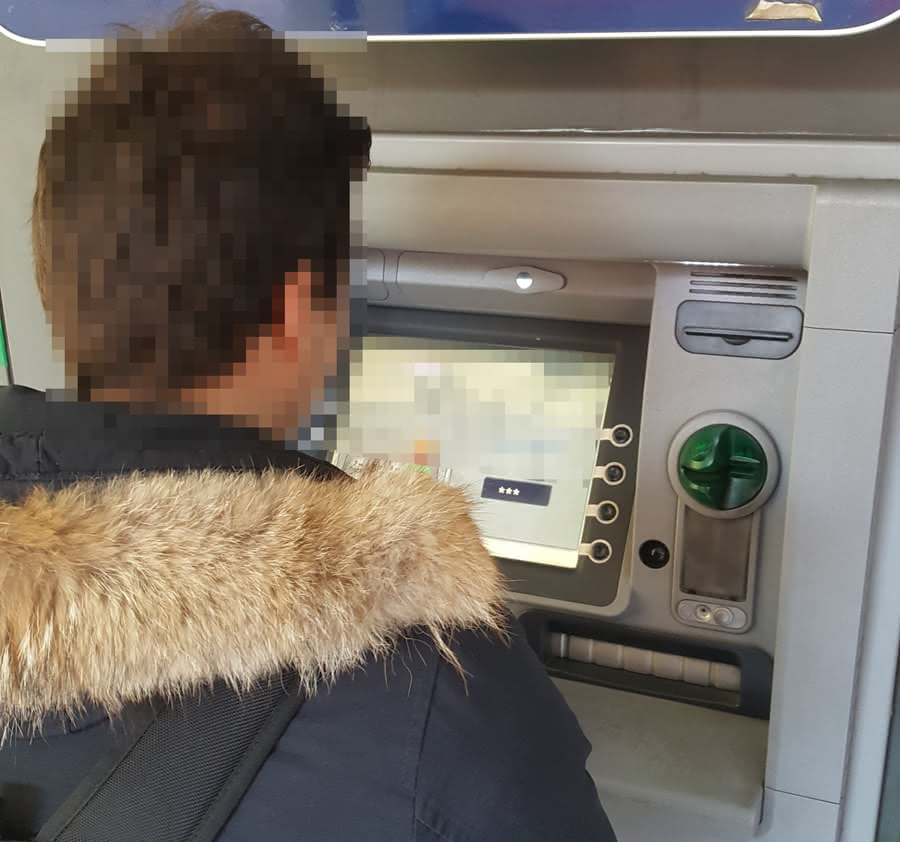}}~
		\subfloat[]{\includegraphics[width=0.48\linewidth]{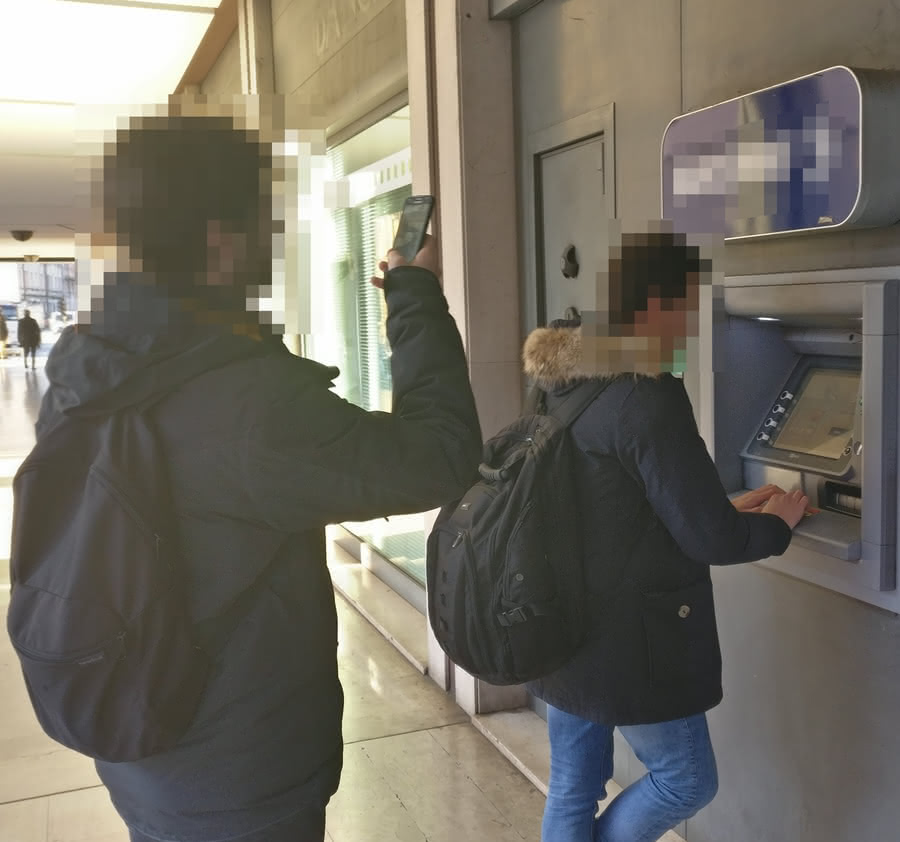}}
		\caption{Attack example -- ATM setting. (a) Adversary's perspective. (b) Outsider's perspective.}
		\label{fig:atm_victim}
	\end{minipage}
\end{figure}

We model a user logging in (authenticating) to a computer system or an ATM using
a PIN or a password ({\em secret}) entered via keyboard or keypad ({\em input
device}). The user receives immediate feedback about each key-press from a
screen, a projector, or both ({\em output device}) in the form of dots or
asterisks ({\em masking symbols}). Shape and/or location of each masking symbol
does not depend on which key is pressed. The adversary can observe and record
the output device(s), though not the input device or the user's hands. An
example of this scenario is shown in Figure \ref{fig:shouldersurfing}. The
adversary's goal is to learn the user's secret.

The envisaged attack setting is representative of many real-world scenarios that
involve low-privilege adversaries, including: \icstart \ic a presenter in a
lecture or conference who types a password while the screen is displayed on a
projector. The entire audience can see the timing of appearance of masking
symbols, and the adversary can be anyone in the audience; \ic an ATM customer
typing a PIN. The adversary who stands in line behind the user might have an
unobstructed view of the screen, and the timing of appearance of masking symbols
(see Figure~\ref{fig:atm_victim}); and \ic a customer enters her debit card PIN at
a self-service gas-station pump. In this case, the adversary can be anyone in
the surroundings with a clear view of the pump's screen.

Although these scenarios seem to imply that adversary is located near the user,
proximity is not a requirement for our attack. For instance, the adversary could
watch a prior recording of the lecture in scenario (1); or, could be monitoring
the ATM machine using a CCTV camera in (2); or, remotely view the screen in (3)
through a compromised IoT camera.
 
Also, we assume that, in many cases, the attack involves multiple observations.
For example, in scenario~(1), the adversary can observe the presenter during
multiple talks, without the presenter changing passwords between talks.
Similarly, in scenario~(2), customers often return to the same ATM.

\section{Overview and Data Collection}\label{sec:atkname}
We assume that the
adversary can capture only inter-keystroke timings leaked by the output device
while the user types a secret. The goal is to analyze differences between the
distribution of inter-keystroke timings and infer corresponding keypairs. This
data is used to identify the passwords that are most likely to be correct, thus
restricting the brute-force search space of the secret. To accurately extract
inter-keystroke timing information, we analyze video feeds of masking symbols,
and identify the frame where each masking symbol first appears. In this setting,
accuracy and resolution of inter-keystroke timings depends on two key factors:
refresh frequency of the output device, and frame rate of the video camera.
Inter-keystroke timings are then fed to a classifier, where classes of interest
are keypairs. Since we assume that the adversary has no access to user-specific
keystroke information, the classifier is trained on population data, rather than
on user-specific timings.

In the rest of this section, we detail the data collection process. We collected
password data from two types of output devices: a VGA-based external projector,
and LCD screens of several laptop computers. See
Section~\ref{sec:password_collection} for details of these devices and corresponding
procedures. For PIN data, we video-recorded the screen of a simulated ATM.
Details can be found in Section~\ref{sec:pins_collection}.

\subsection{Passwords}\label{sec:password_collection}
We collected data using an EPSON EMP-765 projector, and using the LCD screens of
the subjects' laptops computers. In the projector setting, we asked the subjects
to connect their own laptops so they would be using a familiar keyboard. The
refresh rate of both laptop and projector screens were set to 60 Hz -- the
default setting for most systems. This setting introduces quantization errors of
up to about $1/60$ s $\approx16.7$ ms. Thus, events happening within the same
refresh window of $16.7$ms are indistinguishable. We recorded videos of the
screen and the projector using the rear-facing camera of two smartphones:
Samsung Galaxy S5 and iPhone 7 Plus. With both phones, we recorded videos at
$120$ frames per second, i.e., $1$ frame every $8.3$ ms. To ease data
collection, we placed the smartphones on a tripod. When recording the projector,
the tripod was placed on a table, filming from a height of about $165$ cm, to be
horizontally aligned with respect to the projected image. When recording laptop
screens, we placed the smartphone above and to the side of the subject, in order
to mimic the adversary sitting behind the subject.

All experiments took place indoors, in labs and lecture halls at the authors'
institutions. We recruited a total of $62$ subjects, primarily from the student
population of two large universities. Most participants were males in their 20s,
with a technical background and good typing skills. We briefed each subject on
the nature of the experiment, and asked them to type four alphanumerical
passwords:
%
``\texttt{jillie02}'', ``\texttt{william1}'', ``\texttt{123brian}'', and 
``\texttt{lamondre}''.
%
We selected these passwords uniformly at random from the RockYou
dataset~\cite{rockyou_dataset} in order to simulate realistic passwords. The
subjects typed each password three times, while our data collection software
recorded ground-truth keystroke timings of correctly typed passwords with
millisecond accuracy. Timings from passwords that were typed incorrectly were
discarded, and subjects were prompted to re-type the password whenever a mistake
was made. The typing procedure lasted between $1$ and $2$ minutes, depending on
the subject's typing skills. All subjects typed with the ``touch typing'' technique, i.e., using fingers from both hands.

\subsection{PINs}\label{sec:pins_collection}
\begin{figure}[h]
  \centering
  \begin{minipage}{.45\textwidth}
  	\centering
  	\includegraphics[width=.99\linewidth]{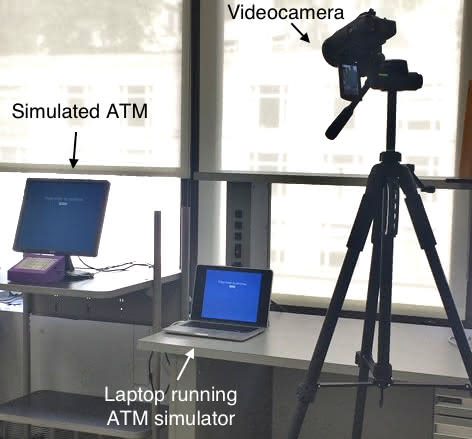}
  	\caption{Setup used in PIN inference experiments.}
  	\label{fig:setup}
  \end{minipage}~~
  \begin{minipage}{.45\textwidth}
	\subfloat[] {\label{fig:simulator_a}
  		\includegraphics[width=0.46\linewidth]{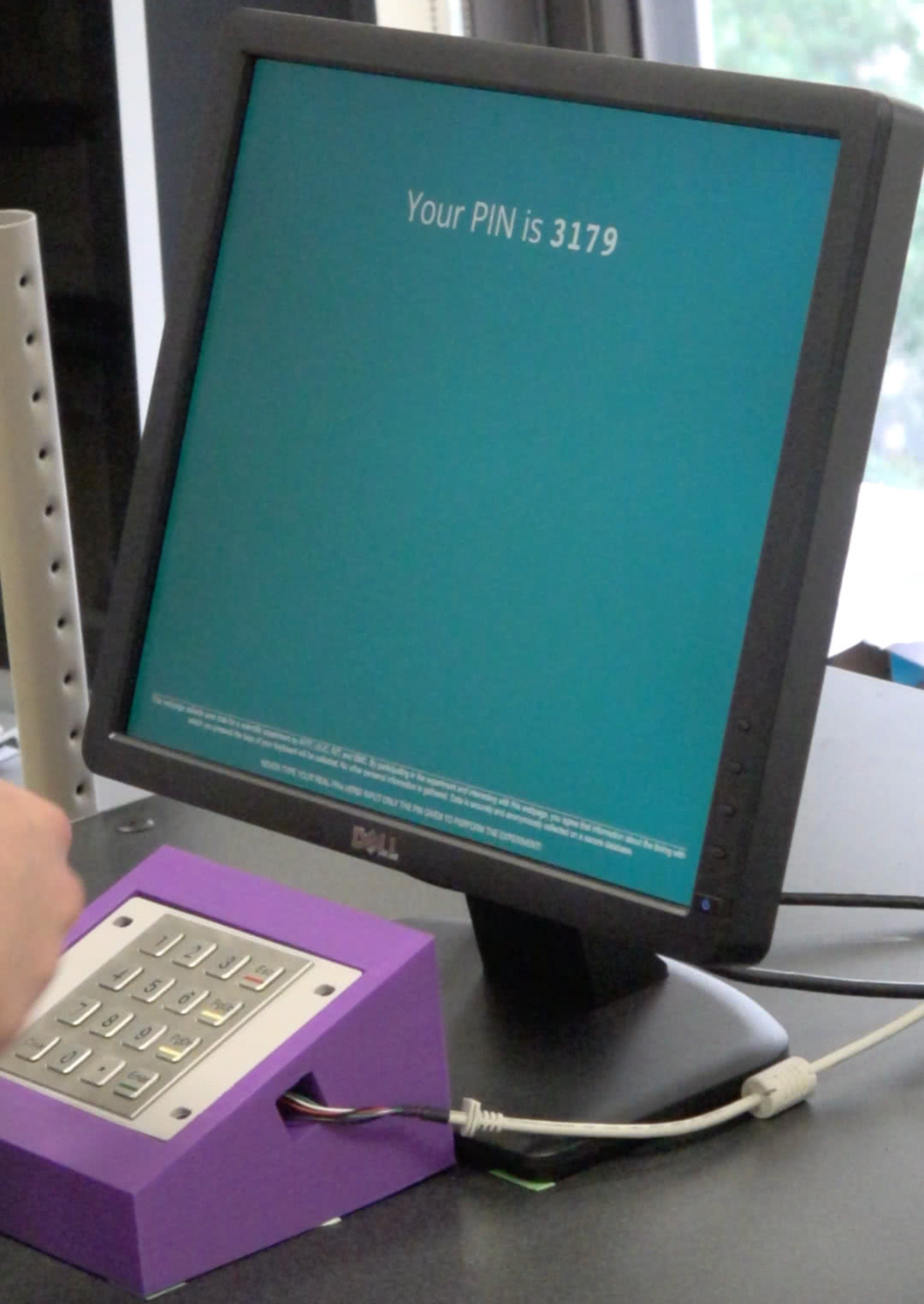}
	}~~~~
	\subfloat[] {\label{fig:simulator_b}
  		\includegraphics[width=0.46\linewidth]{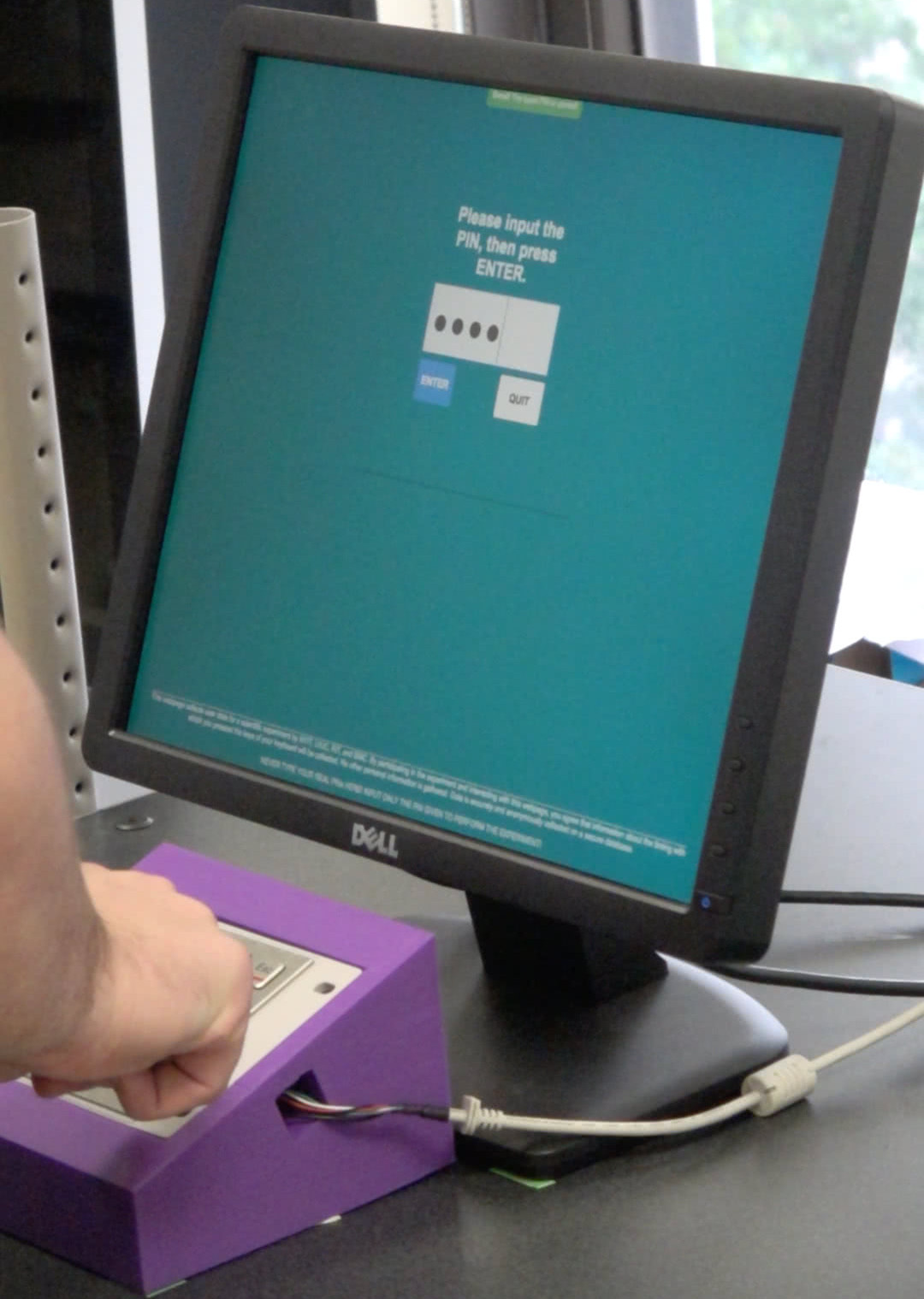}
	}
  	\caption{ATM Simulator during a data collection session. (a) The simulator displays the next PIN. (b) A subject types the PIN from memory.}
  	\label{fig:simulator}
  \end{minipage}\vspace{-0.5cm}
\end{figure}

We recorded subjects entering 4-digit PINs on a simulated ATM, shown in
Figure \ref{fig:setup}. Our dataset was based on experiments with $22$
participants; $19$ subjects completed three data collection sessions, while $4$
subjects completed only one session, resulting in a total of $61$ sessions. At
the beginning of each session, the subject was given $45$ seconds to get
accustomed with the keypad of the ATM simulator. During this time, they were
free to type as they pleased. Next, a subject was shown a PIN on the screen for
ten seconds (Figure~\ref{fig:simulator_a}), and, once it disappeared from the
screen, asked to enter it four times (Figure~\ref{fig:simulator_b}). Subjects
were advised not to read the PINs out loud. 
This process was repeated for $15$ consecutive PINs. During each session,
subjects were presented with the same $15$-PIN sequence $3$ times. Subjects were
given a $30$-second break at the end of each sequence.

\begin{figure}[b]
	\begin{minipage}{.48\textwidth}
		\centering
		\subfloat[] {
			\includegraphics[width=0.44\linewidth]{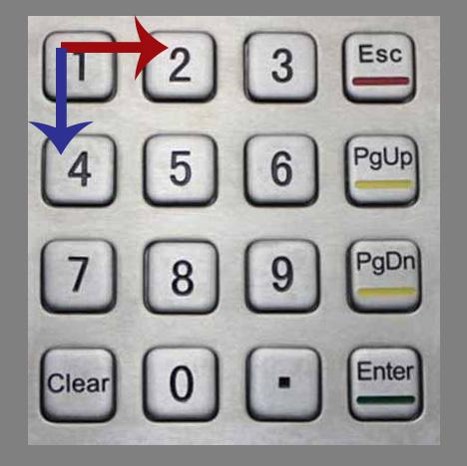}
		}~~
		\subfloat[] {
			\includegraphics[width=0.44\linewidth]{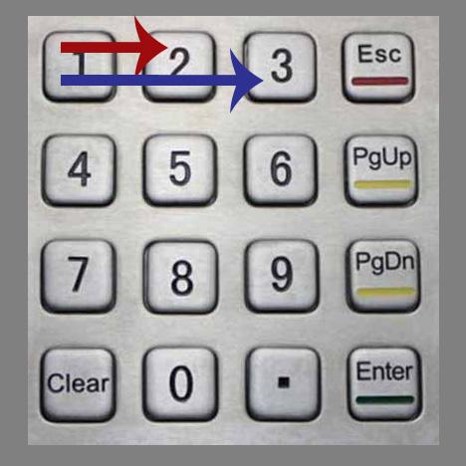}
		}
		\caption{ATM keypad in our experiments. (a) To type keypairs 1-2 and 1-4, the typing finger travels the same distance  in 	different directions. (b) Keypairs 1-2 and 1-3 require the typing finger to travel different distances in the same direction.}
		\label{fig:keypad}
	\end{minipage}~~
	\begin{minipage}{.48\textwidth}
		  \centering
		\includegraphics[width=\linewidth]{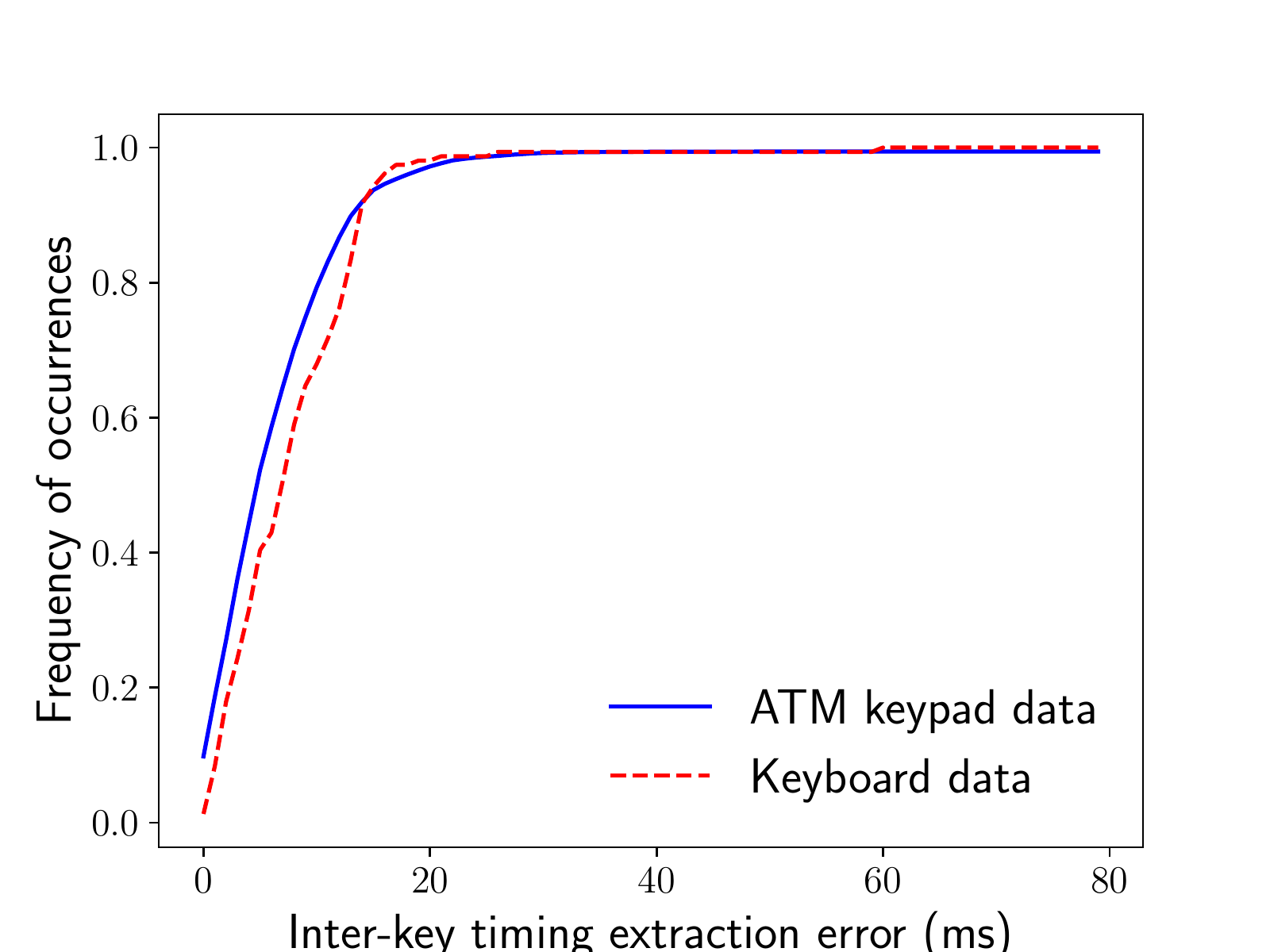}
		\caption{CDF showing error distribution of inter-keystroke timings extracted from videos.} 
		\label{fig:discrepancies}
	\end{minipage}
\end{figure}

Specific 4-digit PINs were selected to test whether: \icstart \ic inter-keypress
time is proportional to Euclidean Distance between keys on the keypad; and \ic
the {\em direction of movement} (up, down, left, or right) between consecutive
keys in a keypair impacts the corresponding inter-key time. We show an example
of these two situations on the ATM keypad in Figure \ref{fig:keypad}. We chose a
set of PINs that allowed collection of a significant number of key combinations
appropriate for testing both hypotheses. For instance, PIN {\tt 3179} tested
horizontal and vertical distance two, while {\tt 1112} tested distance 0 and
horizontal distance 1.

Sessions were recorded using a Sony FDR-AX53 camera, with the pixel resolution
of 1,920$\times$1,080 pixels, and $120$ frames per second. At the same time, ATM
simulation software collected millisecond-accurate inter-key distance ground
truth by logging each keypress. PIN feedback was shown on a DELL $17$" LCD
screen with a refresh rate of $60$ Hz, which resulted to each frame being shown
for $16.7$ ms.

\subsection{Timing Extraction from Video}
We developed software that analyzes video recordings to automatically detect
appearance of masking symbols and log corresponding timestamps. This software
uses OpenCV~\cite{pulli2012real} to infer the number of symbols present in each
image. All frames are first converted to grayscale, and then processed through a
bilateral filter~\cite{tomasi1998bilateral} to reduce noise due to the camera's
sensor. Resulting images are analyzed using Canny Edge
detection~\cite{ding2001canny} to capture the edges of the masking symbol.
External contours are compared with the expected shape of the masking symbol.
When a masking symbol is detected, software logs the corresponding frame number.

Our experiments show that this technique leads to fairly accurate
inter-keystroke timing information. We observed average discrepancy of $8.7$ ms
({\em stdev} of $26.6$ ms) between the inter-keystroke timings extracted from
the video and ground truth recorded by the ATM simulator. Furthermore, 75\% of
inter-keystroke timings extracted by the software had errors under $10$ ms, and
$97$\% had errors under $20$ ms. Similar statistics hold for data recorded on
keyboards for the passwords setting. Figure \ref{fig:discrepancies} shows the
distribution of error discrepancies.

\section{Password Guessing}\label{sec:attack_passwords}
\atkname{} treats identifying digraphs from keystroke timings as a multi-class classification problem, 
where each class represents one digraph, and input to the classifier is a set of inter-keystroke times. 
Without loss of generality, in this section, we assume that the user's password is a sequence of 
lowercase alphanumeric characters typed on a keyboard with a standard layout. 

To reconstruct passwords, we compared two classifiers: Random Forest
(RF)~\cite{ho1995random} and Neural Networks
(NN)~\cite{schalkoff1997artificial}. RF is a well-known classification technique
that performs well for authentication based on keystroke
timings~\cite{bartlow2006evaluating}. Input to RF is one inter-keystroke timing,
and its output is a list of $N$ digraphs ranked based on the probability of
corresponding to input timing. NN is a more complex architecture designed to
automatically determine and extract complex features from the input
distribution. In our experiments, the input to NN is a list of inter-keystroke
timings corresponding to a password. This enables NN to extract features, such
as arbitrary $n$-grams, or timings corresponding to non-consecutive characters.
NN's output is a guess for the entire password. 

\noindent
We instantiated NN using the
following parameters:
\begin{compactitem}
\item number of units in the hidden layer -- $128$ (with ReLU activation functions);
\item inclusion probability of the dropout layer -- $0.2$;
\item number of input neurons -- $25$;
\item number of output layers -- $25$ which represents one character in one-hot encoding. Output layers use softmax activation function;
\item training was performed using batch sizes of $40$ and $100$ epochs. We used the Adam 
optimizer with a learning rate of $0.001$.
\end{compactitem}

\paragraph{Classifier Training.}\label{sec:pwd_training}
We trained \atkname{} on three public datasets~\cite{vural2014shared,roth2014on,banerjee2014emnlp} that contain keystroke timing information
collected from English free-text. Using these datasets for training, we modeled an attack that
relies exclusively on {\em population data}. \new{Without loss of generality, we filtered the datasets to remove
all timings that do not correspond to digraphs composed of alphanumeric
lowercase characters. This is motivated by the datasets' limited availability of digraph samples that contain special characters. In practice, the adversary could collect these timings using, for instance, crowdsourcing tools such as Amazon Mechanical Turk.} To take care of uneven frequencies of different digraphs, we under-represented the most frequent digraphs in the dataset. Data in public datasets was often
gathered from free-text typing of volunteers. Therefore, more frequent digraphs
in English were represented more than rarer ones. For example, considering
\texttt{lamondre}, digraph \texttt{re} appears 43,606 times in the population
dataset, while \texttt{am} -- only 6,481. Similarly, in \texttt{123brian},
digraph \texttt{ri} occurs 19,782 times, while \texttt{3b} -- only 138. We
therefore under-sampled each digraph appearing more than 1,000 times to
1,000 randomly selected occurrences. Similarly, we excluded infrequent digraphs that
appeared under $100$ times in the whole dataset.

\paragraph{Attack Process.}\label{sec:pwd_attack}
To infer the user's secret from inter-keystroke timings, \atkname{}
leverages a dictionary of passwords (e.g., a list of passwords leaked by online
services~\cite{rockyou_dataset,yahoo_leak,pwproject,linkedin_dataset}), possibly
expanded using techniques such as probabilistic context-free
grammars~\cite{weir2009password} and generative adversarial
networks~\cite{hitaj2017passgan}. When evaluating \atkname{}, we assume that the
user's secret is in the dictionary. In practice, this is often the case, as many users use the same weak passwords (e.g., only 36\% of the password of RockYou is unique~\cite{ma2014study}), and reuse them across many different services~\cite{wang2018next,florencio2007large}.
Given that the
size of a reasonable password dictionary is on the order of billions of
entries,\footnote{See for example the lists maintained by
\url{https://haveibeenpwned.com/}.} the goal of \atkname{} is to narrow down the
possible passwords to a small(er) list, e.g., to perform online attacks. This
list is then ranked by the probability associated with each entry, computed from
inter-keystroke timing data. 

Specifically:
\begin{enumerate}
	\item Using RF, for each inter-key time extracted from video
	(corresponding to a digraph), \atkname{} returns a list of $N$ possible
	guesses, sorted by the classifier's confidence. Next, \atkname{} ranks
	the passwords in the dictionary by resulting probabilities as follows:
	for each password, \atkname{} identifies the position in the ranked list
	of predictions for the first digraph of the password being guessed, and
	assigns that position as a ``penalty'' to the password. By performing
	these steps for each digraph, \atkname{} obtains a total penalty score
	for each password, i.e., a score that indicates the probability of the
	password given the output of the RF.
	
	For example, to rank the password \texttt{jillie02}, \atkname{} first
	considers the digraph \texttt{ji}, and the list of predictions of RF for
	the first digraph. It notes that \texttt{ji} appears in such list as the
	$X$-th most probable; therefore, it assigns $X$ as the penalty for
	\texttt{jillie02}. Then, it considers \texttt{il}, which appears in
	$Y$-th position in the list of predictions for the second digraph.
	Penalty for \texttt{jillie02} is thus updated to $X+Y$. This operation
	is repeated for all the $7$ digraphs, thus obtaining the final penalty
	score.
	\item Using NN, \atkname{} computes a list of $N$ possible guesses,
	sorted by the classifier's confidence of each guess. In this case, the
	\atkname{} processes the entire list of flight times at once, rather
	than refining its guess with each digraph.
\end{enumerate}
We considered the following attack settings: {\em single-shot}, and {\em
multiple recordings}. With the former, the adversary trains \atkname{} with
inter-keystroke timings from population data, i.e., from users other than the
target, e.g., from publicly available datasets, or by recruiting users and
asking them to type passwords. In this scenario, the adversary has access to the
video recording of a single password entry session. With multiple recordings, the
adversary trains \atkname{} as before, and additionally, has access to videos of
multiple login instances by the same user.

\new{Training \atkname{} exclusively with population data leads to
more realistic attack scenarios than training it with user-specific data, because usually the adversary has limited access to keystrokes samples
from the target user. Further, access to user-specific data will likely improve the success rate of \atkname{}.
}

\subsection{Results}\label{sec:pwd_results}
In this section, we report on \atkname{} efficacy in reducing search time on the
RockYou~\cite{rockyou_dataset} password dataset compared to random choice,
weighted by probability. We restricted experiments to the subset of
$8$-character passwords from RockYou, since the adversary can always determine
password length by counting the number of masking symbols shown on the screen.
This resulted in 6,514,177 passwords, out of which 2,967,116 were unique.

\paragraph{Attack Baseline.}
\new{To establish the attack baseline, we consider an adversary that outputs password guesses from a leaked dataset in descending order of frequency. (Ties are broken using random selection from the candidate passwords.)
Because password probabilities are far from uniform 
(e.g., in RockYou, top $200$ $8$-character passwords account for over $10$\% of 
the entire dataset), this is the best adversarial strategy given no additional information on the target user. 
}

Passwords selected for our evaluation represent a mix of common and rare
passwords. Thus, they have widely varying frequencies of occurrence in RockYou
and expected number of attempts needed to guess each password using the baseline
attack varies significantly. For example, expected number of attempts for:
\begin{compactitem}
\item \texttt{123brian} (appears $6$ times) -- 93,874;
\item \texttt{jillie02}, (appears only once) -- 1,753,571;
\item \texttt{lamondre} (appears twice) -- 397,213;
\item \texttt{william1}  (appears 1,164 times) -- only 187.
\end{compactitem}

\begin{figure}[t]
	\centering
	\includegraphics[width=.6\linewidth]{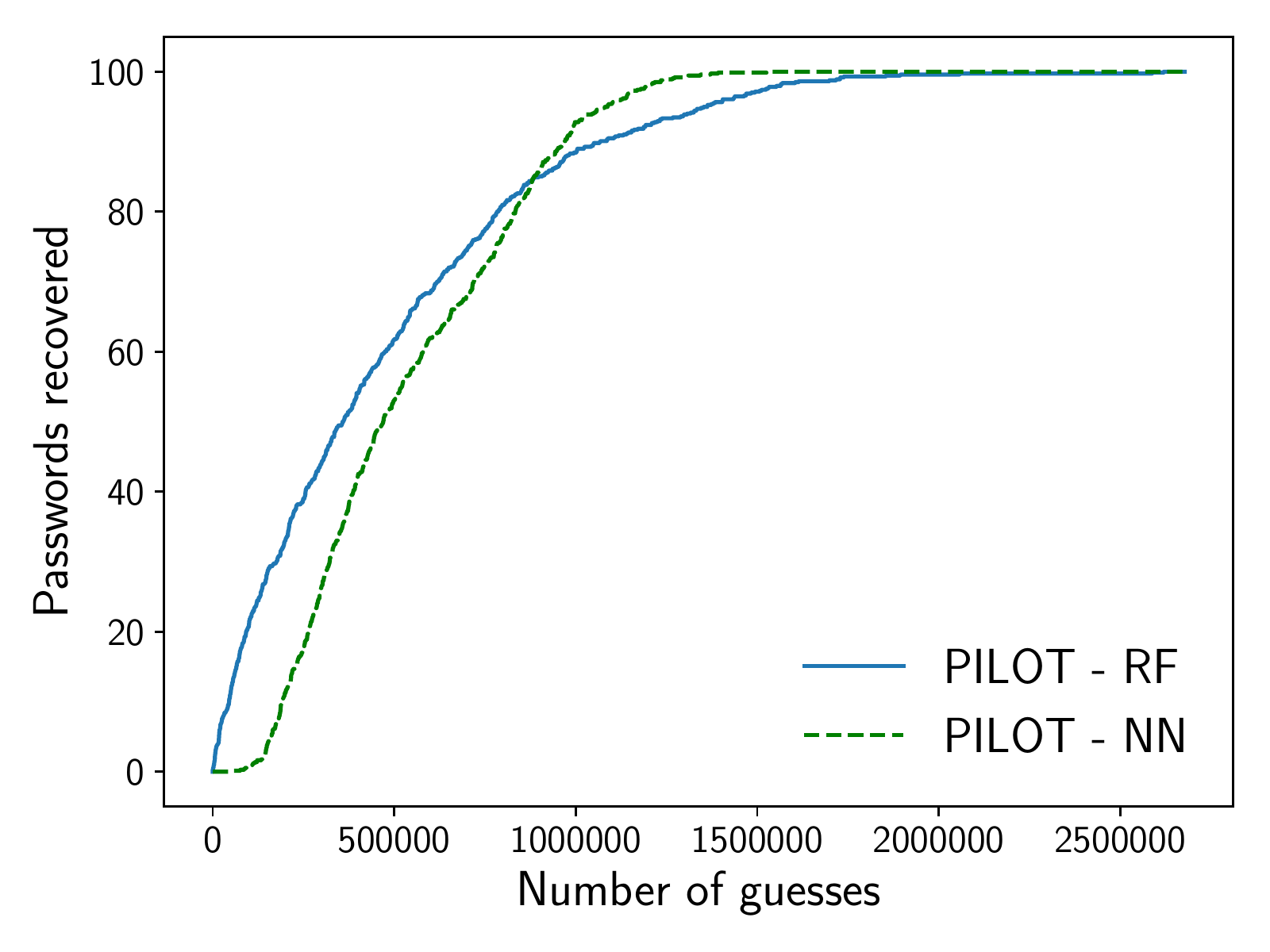}
	\caption{CDF of the amount of passwords recovered by \atkname{}---\textit{Population Data} attack scenario.}
	\label{fig:password_population_all}
\end{figure}

\begin{figure*}[t]
	\centering
	\subfloat[\texttt{123brian} (183 auth. attempts).]{\includegraphics[width=.5\textwidth]{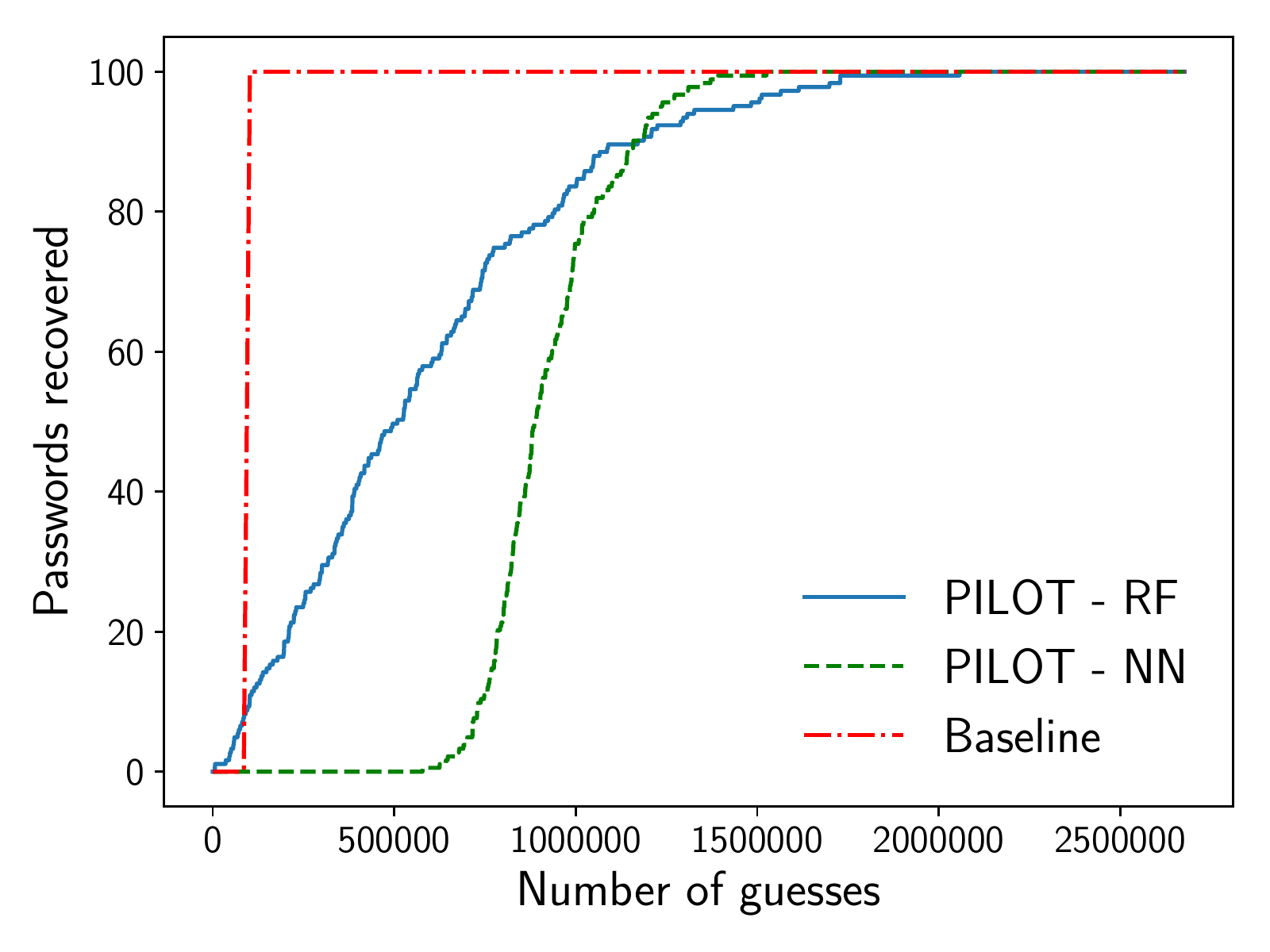}\label{subfig:passwor_population_123brian}}%
	\subfloat[\texttt{jillie02} (186 auth. attempts).]{\includegraphics[width=.5\textwidth]{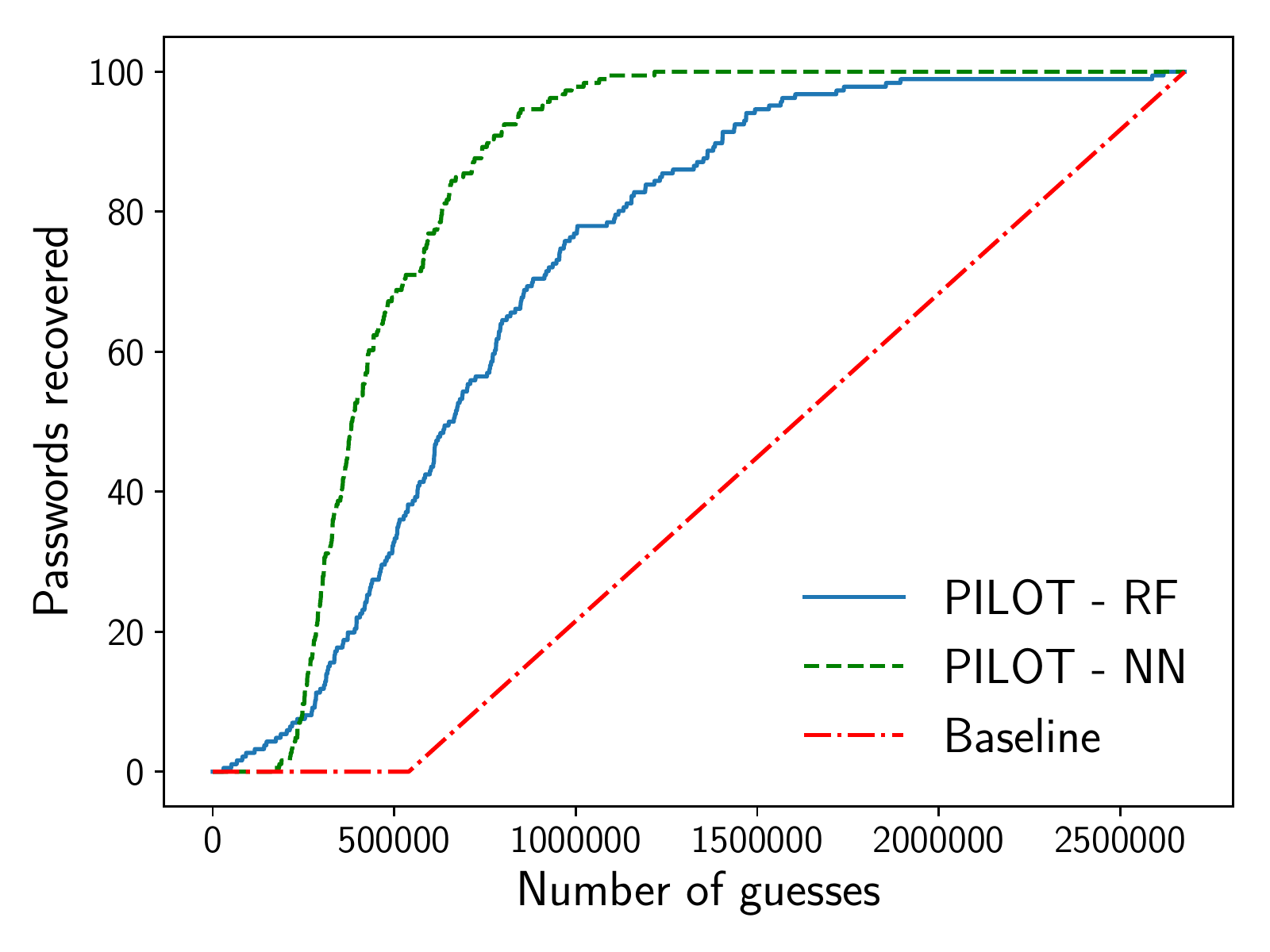}}\\
	\subfloat[\texttt{lamondre} (184 auth. attempts).]{\includegraphics[width=.5\textwidth]{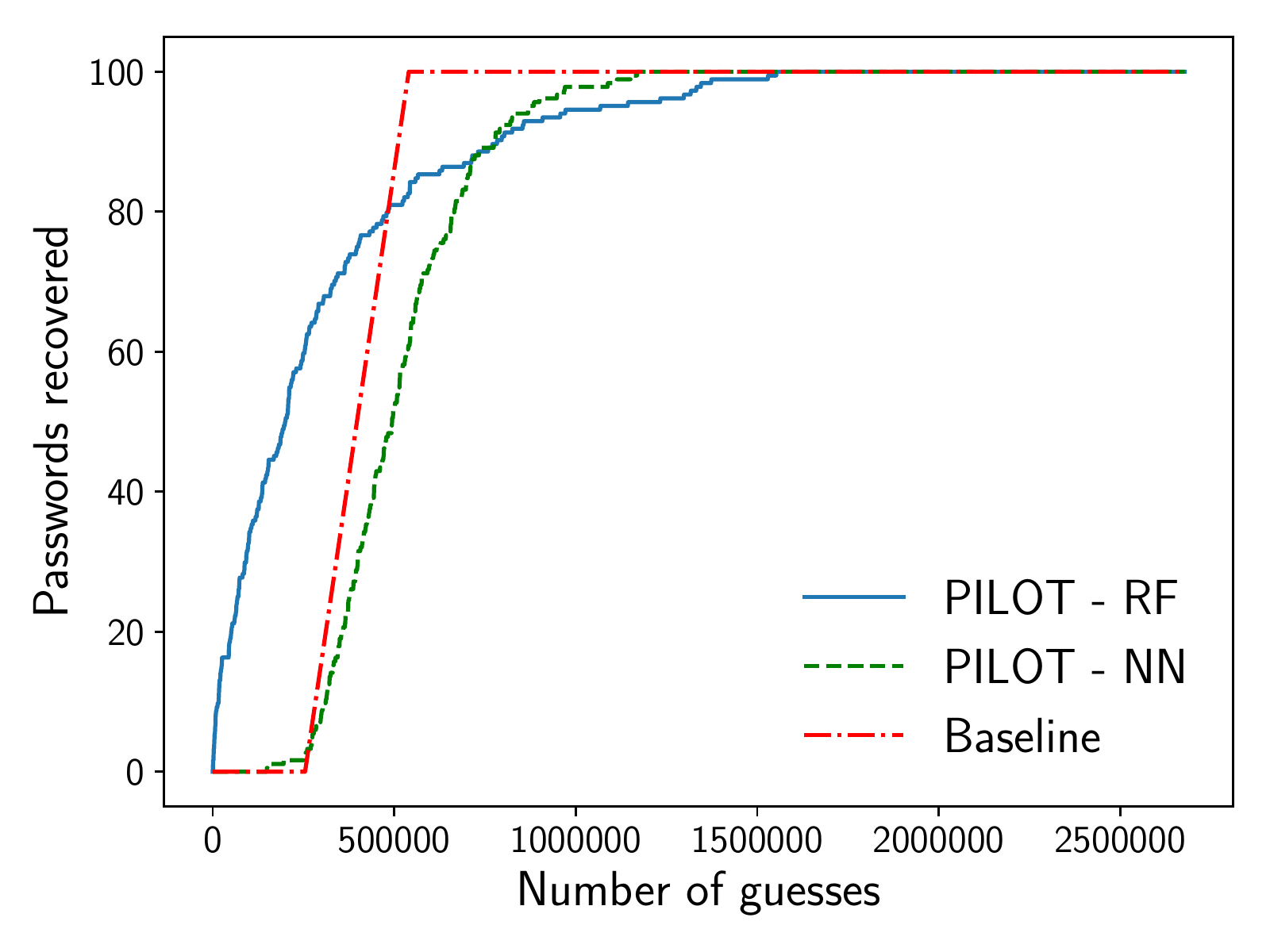}}
	\subfloat[\texttt{william1} (183 auth. attempts).]{\includegraphics[width=.5\textwidth]{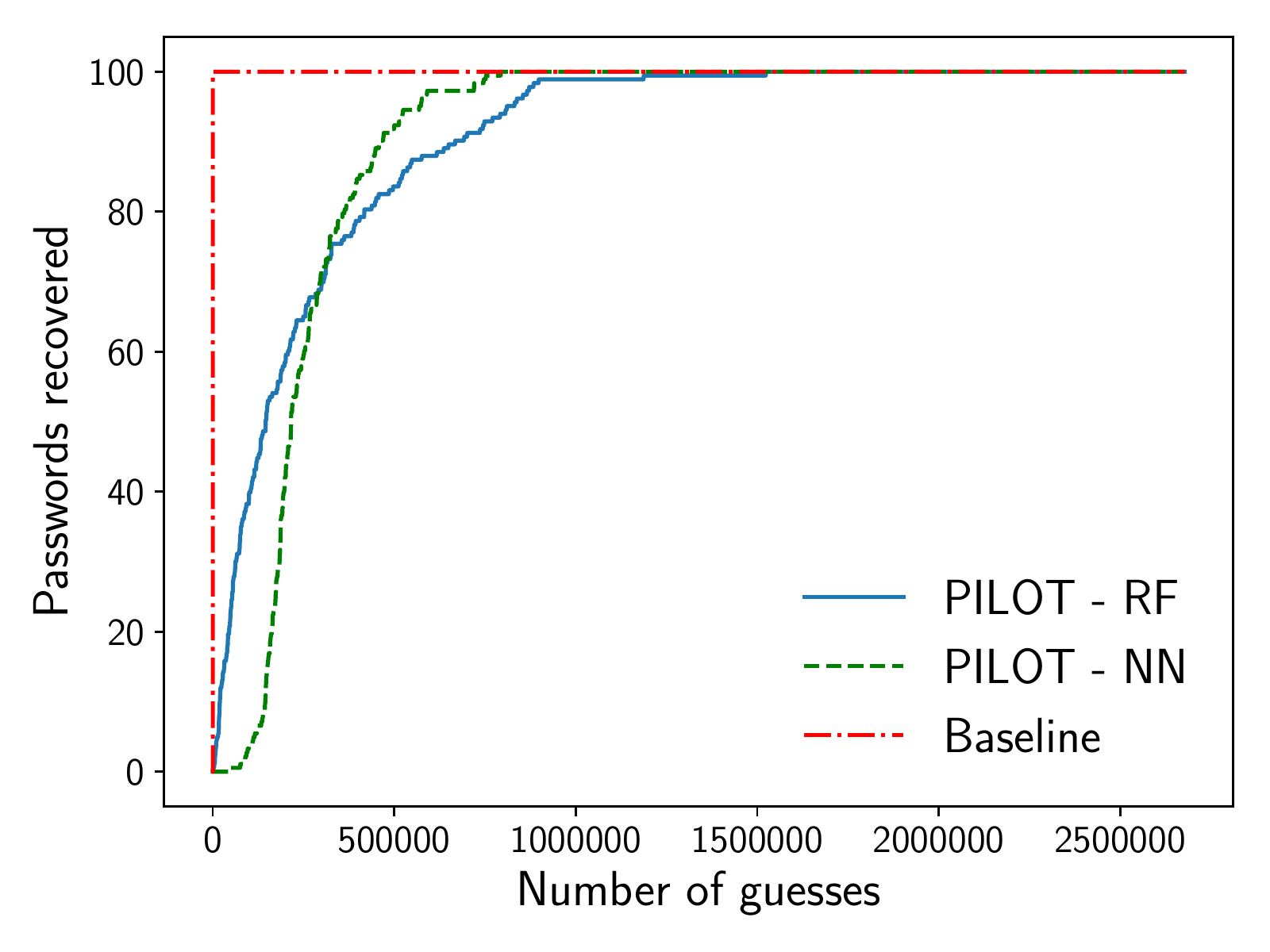}}
	\caption{CDF for the number of passwords recovered by \atkname{}, for each target password. 
	Plots also show the baseline attack for the corresponding password.\vspace{-0.5cm}}
	\label{fig:password_population_sep}
\end{figure*}

\paragraph{Single-shot.} 
Results in the single-shot setting are summarized in
Table~\ref{tab:password_population}. Cumulative Distribution Function (CDF) of
successfully recovered passwords is reflected in
Figure \ref{fig:password_population_all}, and breakdown of results (by target
password) is shown in Figure \ref{fig:password_population_sep}. 

{
	\tabulinesep = 2mm
	\begin{table*}[!t]
		\caption{\atkname{}---\textit{Single-shot} setting. \textit{Avg}: average number of attempts  
			to guess a password; \textit{Stdev}: standard deviation; \textit{Rnd}: number of guesses for the baseline adversary; \textit{$<$Rnd}: how 
			often \atkname{} outperforms random guessing; \textit{Best}: number of attempts of the best guess; 
			\textit{$<n$}: how many passwords are successfully guessed within 
			first $n=$~20,000/100,000 attempts.} 
		\label{tab:password_population}
		\begin{tabu} to \textwidth {X[0.8]X[0.5r]X[0.5r]X[0.5r]X[0.6r]X[0.45r]X[0.5r]X[0.4r]X[0.4r]}
			\tabucline-
			\rowfont \bfseries & Avg & Stdev & Med & Rnd & $<$Rnd & 
			Best & $<$20k & $<$100k\\
			\hline\hline 
			\multicolumn{9}{c}{\textbf{Random Forest}}\\
			\hline

			\texttt{123brian} & 581,743 & 414,761 & 508,332 & 93,874 & 8.7\% & 5,535 & 1.1\% &9.3\%\\
			\texttt{jillie02} & 749,718 & 448,319 & 656,754 & 1,753,571 & 97.8\% & 28,962 & 0.0\% & 2.7\%\\
			\texttt{lamondre} & 301,906 & 334,681 & 199,344 & 397,213 & 75.0\% & 145 & 13.0\% &  33.7\%\\
			\texttt{william1} & 246,437 & 264,090 & 145,966 & 187 & 0.5\% & 68 & 10.9\% &  39.9\%\\
			\hline\hline 
			\multicolumn{9}{c}{\textbf{Neural Network}}\\
			\hline
			\texttt{123brian} & 923,534 & 165,454 & 886,802 & 93,874 & 0.0\% & 577,739 & 0.0\% & 0.0\% \\
			\texttt{jillie02} & 456,811 & 210,512 & 383,230 & 1,753,571 & 100.0\% & 164,754 & 0.0\% & 0.0\%\\
			\texttt{lamondre} & 517,472 & 189,355 & 493,713 & 397,213 & 28.8\% & 148,403 & 0.0\% & 0.0\%\\
			\texttt{william1} & 265,813 & 140,753 & 215,840 & 187 & 0.0\% & 45,176 & 0.0\% & 3.8\%\\
			\hline 
		\end{tabu}
	\end{table*}
}

Results show that, for uncommon passwords (\texttt{jillie02} and
\texttt{lamondre}), \atkname{} consistently outperforms random guessing. In
particular, for \texttt{jillie02} both RF and NN greatly exceed random guessing,
since both their curves in Figure \ref{fig:password_population_sep} are above
random guess baseline. For \texttt{lamondre}, RF shows an advantage over random
guess in 76\% of the instances, while NN never beats the baseline. 

For common passwords, sorted random guess wins over \atkname{}. In particular,
\texttt{123brian} is both popular (i.e., 93,874-th most popular password of the
set, corresponding to the top $3$\% of the RockYou dataset) and very hard to
recover with \atkname{}. This can be observed from
Figure \ref{fig:password_population_sep}, where the curves corresponding to
\texttt{123brian} are least steep. Finally, \texttt{william1}, being the
$187$-th most popular password, is always recovered early in our baseline attack, with the notable exception of one instance by RF.

In general, \atkname{} wins over the sorted random guess on infrequent
passwords, such as \texttt{jillie02} and \texttt{lamondre}, that appear only
once or twice, respectively. Such infrequent passwords exhibit the same random
guess baseline curve and average, reported in Table~\ref{tab:password_population}
and shown in Figure \ref{fig:password_population_sep}. Given the similar steepness
of CDF curves in Figure \ref{fig:password_population_sep}, which hint that
\atkname{}\ 's performance might be similar for many other passwords, \atkname{}
can probably outperform the baseline for uncommon passwords. We also note that
uncommon passwords represent the vast majority of user-chosen passwords: 90\% of RockYou passwords 
appear at most twice, and 80\% exactly once. 
\new{We expect that a realistic adversary would first generate password guesses based on their frequency alone (as in our baseline attack), and then switch to \atkname{} once these frequencies drop below some threshold.}


Finally, we highlight that random guess baseline is computed on the distribution
of passwords in RockYou. Other datasets might have different distributions: for
example, in the \textit{10 million password list} dataset~\cite{10mpwds},
\texttt{jillie02}, \texttt{lamondre}, and \texttt{123brian} appear only once,
while \texttt{william1} appears $176$ times.

We believe that the discrepancy between performance of \atkname{} on various
passwords is due to how frequently the digraphs in each password appear in
training data. Specifically, even with our under-representation, all digraphs in
\texttt{william1}, with the exception of \texttt{m1}, are far more frequent in
the training data than \texttt{12}, \texttt{23}, \texttt{3b}, or \texttt{02}. 

Regarding specific classifiers, RF overtakes NN in most instances. For example,
when guessing \texttt{123brian} (Figure~\ref{subfig:passwor_population_123brian}),
NN performs worse than random guessing for first 800,000 attempts. Afterwards,
NN outperforms both random guessing and RF. Furthermore, while RF can guess a
substantial percentage of passwords within 20,000, 50,000 and 100,000
attempts, NN cannot achieve the same result. 

In terms of minimum number of guesses per password, RF recovered
\texttt{william1} in 68, \texttt{lamondre} in 145, \texttt{123brian} in
5,535, and \texttt{jillie02} in 28,962 attempts. NN required a consistently
higher minimum number of attempts for each password.

\paragraph{Multiple Recordings.} 
Information from three login instances was used as follows. We averaged classifiers' predictions over three login instances for a given user, and ranked passwords accordingly. 

{
	\tabulinesep = 2mm
	\begin{table*}[!h]
		\caption{\atkname{}---\textit{Multiple recordings} setting. \textit{Avg}: average number of attempts  
			to guess a password; \textit{Stdev}: standard deviation; \textit{Rnd}: number of guesses for the baseline adversary; \textit{$<$Rnd}: how 
			often \atkname{} outperforms random guessing; \textit{Best}: number of attempts of the best guess; 
			\textit{$<n$}: how many passwords are successfully guessed within 
			first $n=$~20,000/100,000 attempts.} 
		\label{tab:password_recurring}
		\begin{tabu} to \textwidth {X[0.8]X[0.5r]X[0.5r]X[0.5r]X[0.6r]X[0.45r]X[0.5r]X[0.4r]X[0.4r]}
	\tabucline-
	\rowfont \bfseries & Avg & Stdev & Med & Rnd & $<$Rnd & 
	Best & $<$20k & $<$100k\\
			\hline\hline 
			\multicolumn{9}{c}{\textbf{Random Forest}}\\
			\hline
			\texttt{123brian} & 552,574 & 468,539 & 402,166 & 93,874 & 14.1\% & 13,931 & 4.7\% & 14.1\%\\
			\texttt{jillie02} & 713,895 & 410,225 & 606,403 & 1,753,571 & 100.0\% & 67,875 & 0.0\% & 1.6\%\\
			\texttt{lamondre} & 398,186 & 425,811 & 236,905 & 397,213 & 65.6\% & 404 & 6.2\% & 25.0\%\\
			\texttt{william1} & 370,933 & 602,654 & 148,405 & 187 & 1.6\% & 19 & 17.2\% & 42.2\%\\
			\hline\hline 
			\multicolumn{9}{c}{\textbf{Neural Network}}\\
			\hline
			\texttt{123brian} & 922,655 & 129,927 & 889,406 & 93,874 & 0.0\% & 676,418 & 0.0\% & 0.0\%\\
			\texttt{jillie02} & 439,414 & 155,385 & 402,332 & 1,753,571 & 100.0\% & 205,645 & 0.0\% & 0.0\%\\
			\texttt{lamondre} & 503,248 & 137,276 & 504,493 & 397,213 & 21.3\% & 182,123 & 0.0\% & 0.0\%\\
			\texttt{william1} & 248,769 & 103,240 & 216,630 & 187 & 0.0\% & 86,213 & 0.0\% & 1.6\%\\
			\tabucline-
			\tabuphantomline
		\end{tabu}
	\end{table*}
}

\begin{figure}[t]
	\centering
		\includegraphics[width=.6\linewidth]{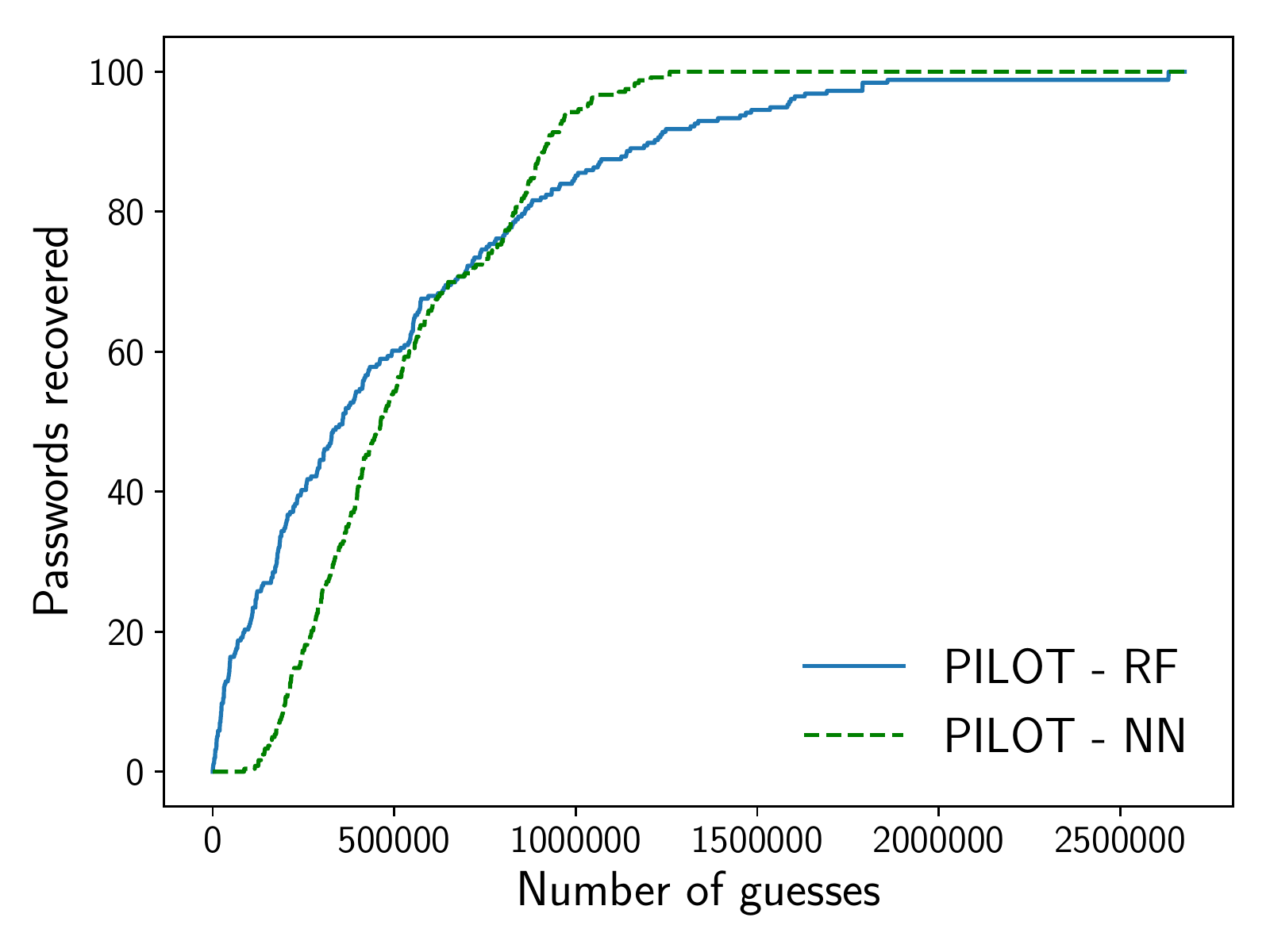}
	\caption{CDF showing number of passwords recovered by \atkname{} in the \textit{Multiple recordings} scenario.}\vspace{-0.5cm}
	\label{fig:password_recurring_all}
\end{figure}

Results are summarized in Table~\ref{tab:password_recurring}, and
Figure \ref{fig:password_recurring_all}. Although \atkname{} still consistently
outperforms random guessing, using data from multiple authentication recordings
leads to mostly identical results overall with both RF and NN. \atkname{}~'s
guessing success rate for \texttt{123brian} and \texttt{jillie02} is slightly
improved compared to the previous setting and minimum number of attempts to
recover each password diminished slightly. We recovered \texttt{william1} in
19, \texttt{lamondre} in 404, \texttt{123brian} in 13,931, and
\texttt{jillie02} in 67,875 attempts. Overall, results show that there are no
substantial benefits in using timing data from three recordings from the same
user.

\section{PIN Guessing}\label{sec:attack_pins}
In this section, we present our results on PIN-guessing. First, we analyze the
relationship between inter-keystroke timings and Euclidean Distance between
consecutive keys, and between inter-keystroke timings and direction of movement
on the keypad. We then show how the adversary can use timing information to infer key distances, and therefore to predict PINs.

We are not aware of any publicly-available PIN timing datasets that can be used
to train \atkname{}. To address this issue, we collected data from 22 users. To
compute the attack baseline, we considered all PINs to be equally likely, i.e.,
we are modeling PINs as random four-digit strings. This is consistent with how
many European banks assign PINs to bank cards~\cite{fineco}, and with the work
of Bonneau et al.~\cite{bonneau2012pin}, which showed that users are reluctant
to change the random PIN provided by their bank.

\begin{figure}[t!]
  \centering
  \subfloat[From raw data.]{
  \includegraphics[width=.48\linewidth]{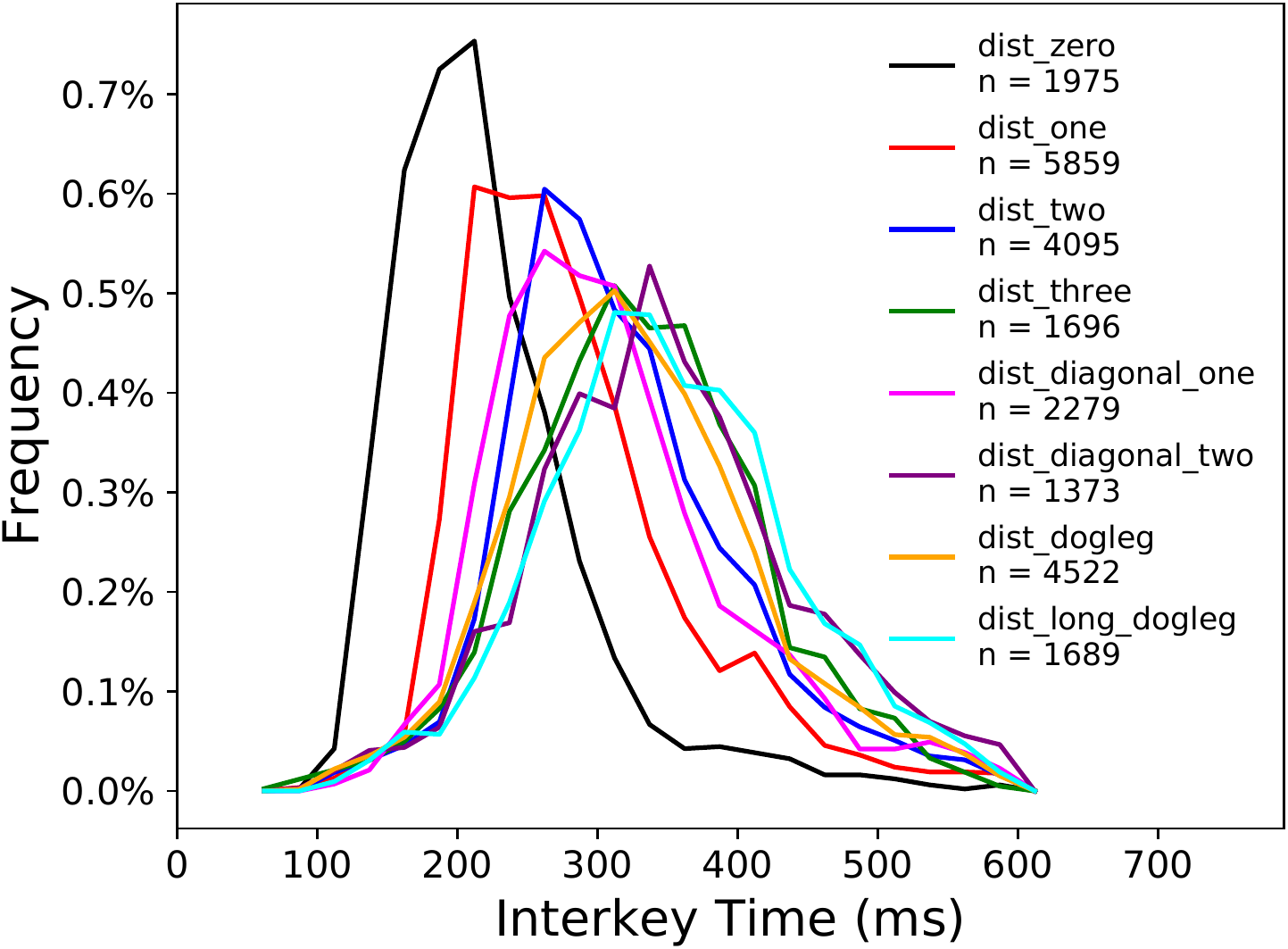}\label{fig:pdf_distances}}~
\subfloat[Modeled as gamma distributions.]{\includegraphics[width=.45\linewidth]{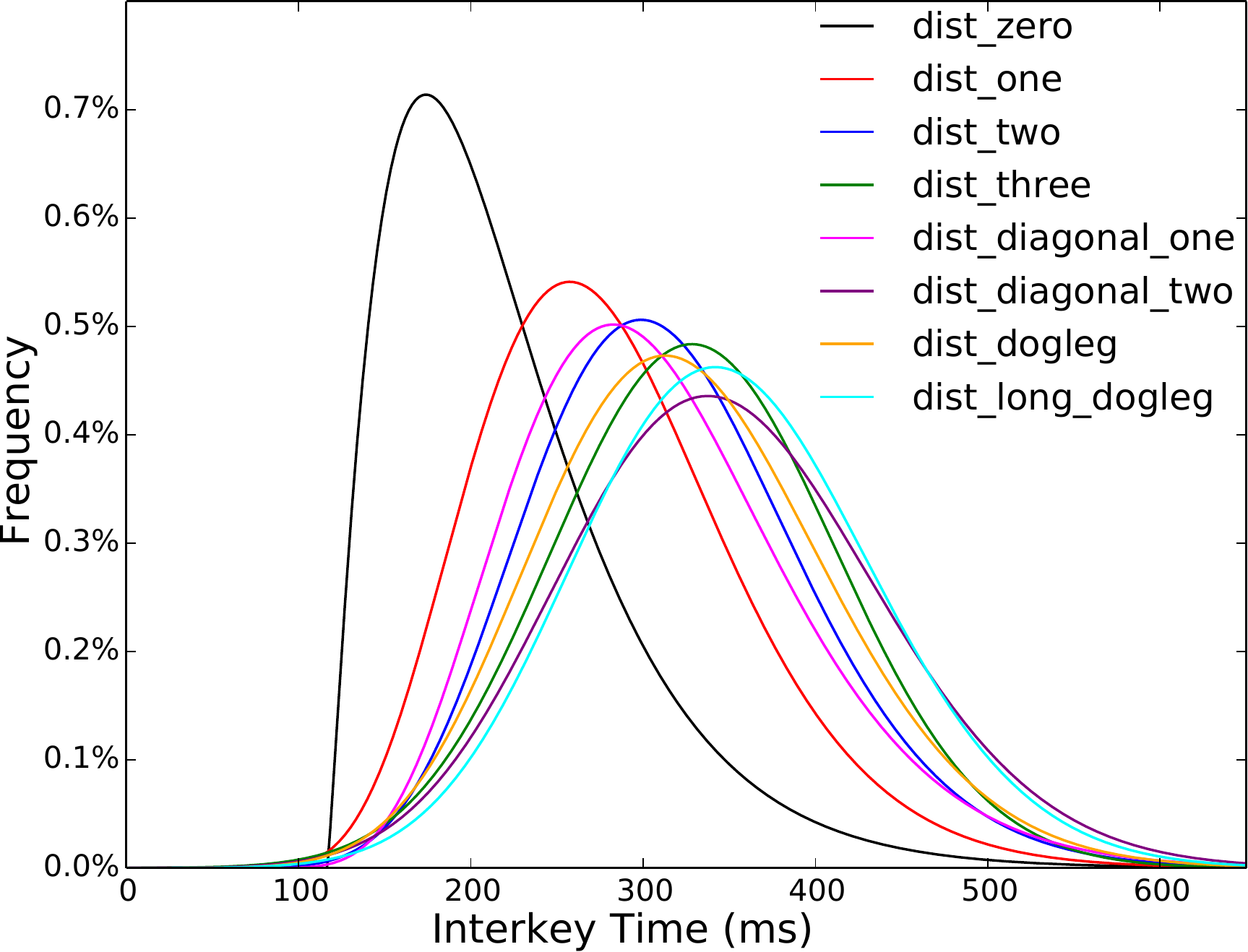}\label{fig:gamma}}
\caption{Inter-keystroke timings of all possible distances for ATM keypad typing.}
\label{smth}\vspace{-0.5cm}
\end{figure}

\paragraph{Distance.}
\label{subsubsec:distance}
Across all subjects, we observed that distributions of inter-keystroke latencies
were distinct in all cases (for $p$-value $< 5\cdot~10^{-6}$), with the
following exceptions: (1) latencies for distance 2 (e.g., keypair 1-3) were
close to latencies for distance 3 (keypair 2-0); (2) latencies for distance 2
were close to latencies for diagonal 1$\times$1 (e.g., keypair 4-8); latencies
for distance 3 were close to latencies for 2$\times$1 diagonal (i.e. ``2'' to
``9'', ``1'' to ``6'', etc.), and diagonal 2$\times$2 (e.g., keypair 7-3), and
diagonal 3$\times$2 (e.g., keypair 3-0). Figure \ref{fig:pdf_distances} shows the
various probability distributions, while Figure \ref{fig:gamma} models these
different probability distribution functions as gamma distributions. In
Figure \ref{fig:pdf_distances}, {\sf dist\_zero} indicates keypairs
composed of the same two digits. {\sf dist\_one}, {\sf dist\_two}, and {\sf
dist\_three} shows timings distributions for keypairs with horizontal or
vertical distance one (e.g., keypair 2-5), two (e.g., 2-8), and three (2-0),
respectively. {\sf dist\_diagonal\_one} and {\sf dist\_diagonal\_two}
indicates keypairs with diagonal distance one (e.g., 2-4) and distance two
(e.g., 1-9), respectively. {\sf dist\_dogleg} and {\sf dist\_long\_dogleg}
show timing distributions of keypairs such as 1-8 and 0-3. In
Figure \ref{fig:gamma}, {\sf dist\_one\_horizontal} and {\sf dist\_one\_vertical}
indicate Euclidean Distance right in the left/right directions, and up/down
directions, respectively, while {\sf dist\_one\_up}, {\sf dist\_one\_down},
{\sf dist\_one\_left}, and {\sf dist\_one\_right} indicate distances one in
the up, down, left, and right directions.
 
\begin{figure}[h]
\begin{minipage}{.48\textwidth}
  \centering
  \includegraphics[width=\linewidth]{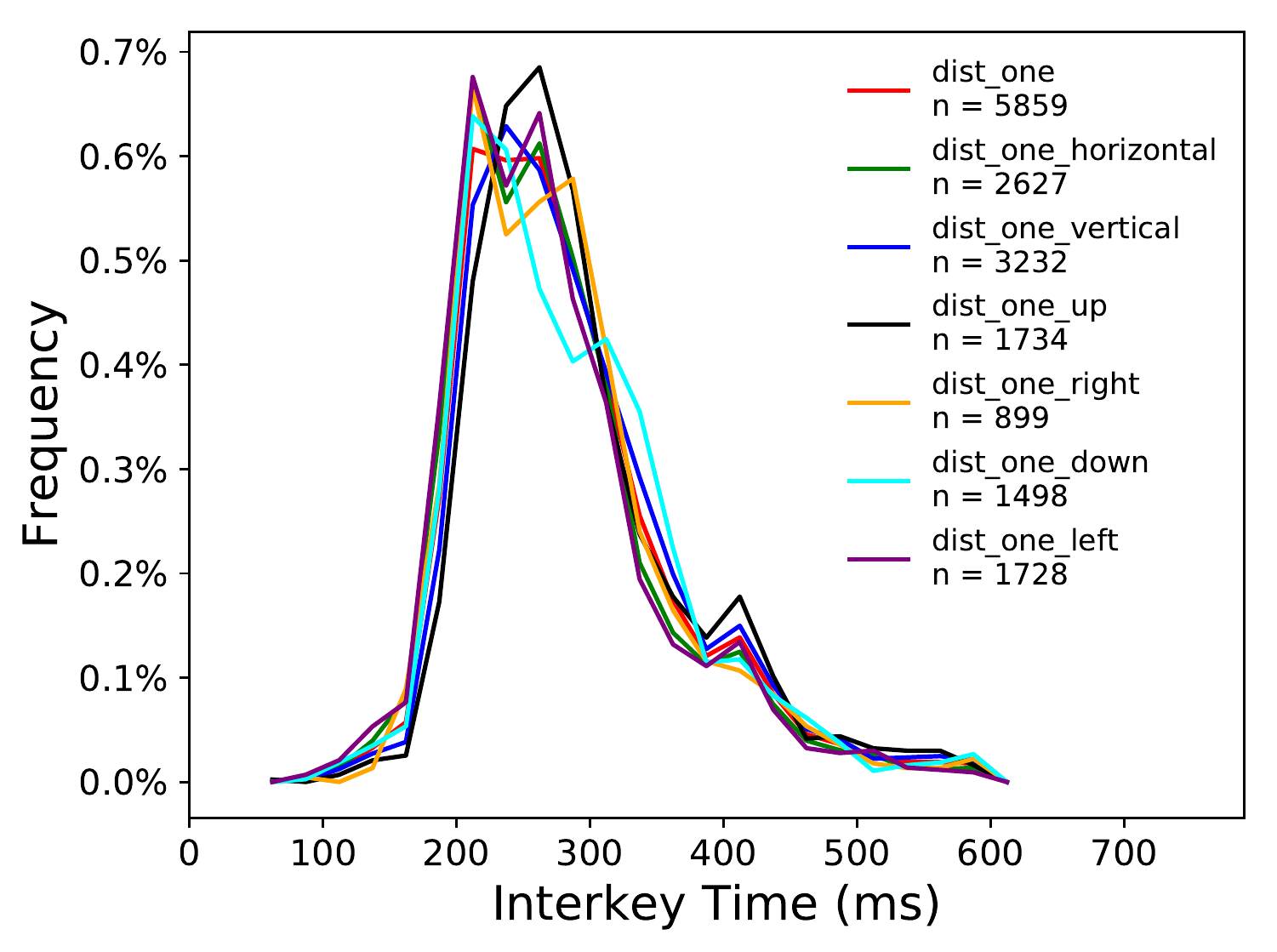}
  \caption{Frequency of inter-keystroke timings for Euclidean Distance of one. {\sf dist\_one} indicates 
  latency distribution for distance one in any direction.}
  \label{fig:pdf_distance1}
\end{minipage}~~
\end{figure}

\paragraph{Direction.}
\label{subsubsec:direction}
The relative orientation of key pairs characterized by the same Euclidean
distance (e.g., 2-3 vs.~2-5) has a negligible impact on the corresponding
inter-key latency. We observed that the distributions of keypress latencies
observed from each possible direction between keys were not significantly
different (for $p$-value $< 10^{-4}$). Figure \ref{fig:pdf_distance1} shows
different probability distributions relative to various directions for Euclidean
distance 1.

\subsection{Pin Inference}
\label{subsec:pininference}

Using the data we collected, we mapped the distribution of inter-keypress
latencies, and used the resulting probabilities to test the effectiveness of
PINs prediction from inter-key latencies.
Our PIN guessing algorithm is composed of two parts: (1) an algorithm that
estimates distances from keystroke timings; and (2) an algorithm that ranks PINs
based on the estimated distances.
%
The core idea is to consider the PIN pad
as a weighed multigraph. The graph nodes represent the keys, and are labeled
0-9. Keys are connected by weighted edges. The weight of an edge corresponds to
the Euclidean distance between the corresponding keys, using the distance
between two adjacent keys (e.g., 1 and 2) as unit. We identified 8 possible
distances: zero distance (e.g., key 3 followed by key 3, $weight = 0$);
horizontal or vertical distance one (e.g. keys 1-2, $weight = 1$); horizontal or
vertical distance two (e.g., keys 1-3, $weight = 2$); vertical distance three
(e.g., keys 2-0, $weight = 3$); diagonal distance one (e.g., keys 1-5, $weight =
\sqrt{1^2 + 1^2}$); diagonal distance two (e.g., keys 1-9, $weight =
\sqrt{2^2+2^2}$); short diagonal distance (e.g., keys 1-8, $weight = \sqrt{1^2 +
2^2}$) and long diagonal distance (e.g. keys 1-0, $weight = \sqrt{1^2 + 3^2}$).

For each PIN, we created three sets a subgraphs, indicated as $S_1$, $S_2$, and $S_3$, composed only of the nodes connected by edges with the same weight as the estimated distance. Specifically, $S_i$ contains all the two-nodes subgraphs such that their edges have weight equal to the estimated distance between the keys in the $i$-th PIN digraph. 

\begin{figure}[t]
	\centering
		\includegraphics[width=1\linewidth]{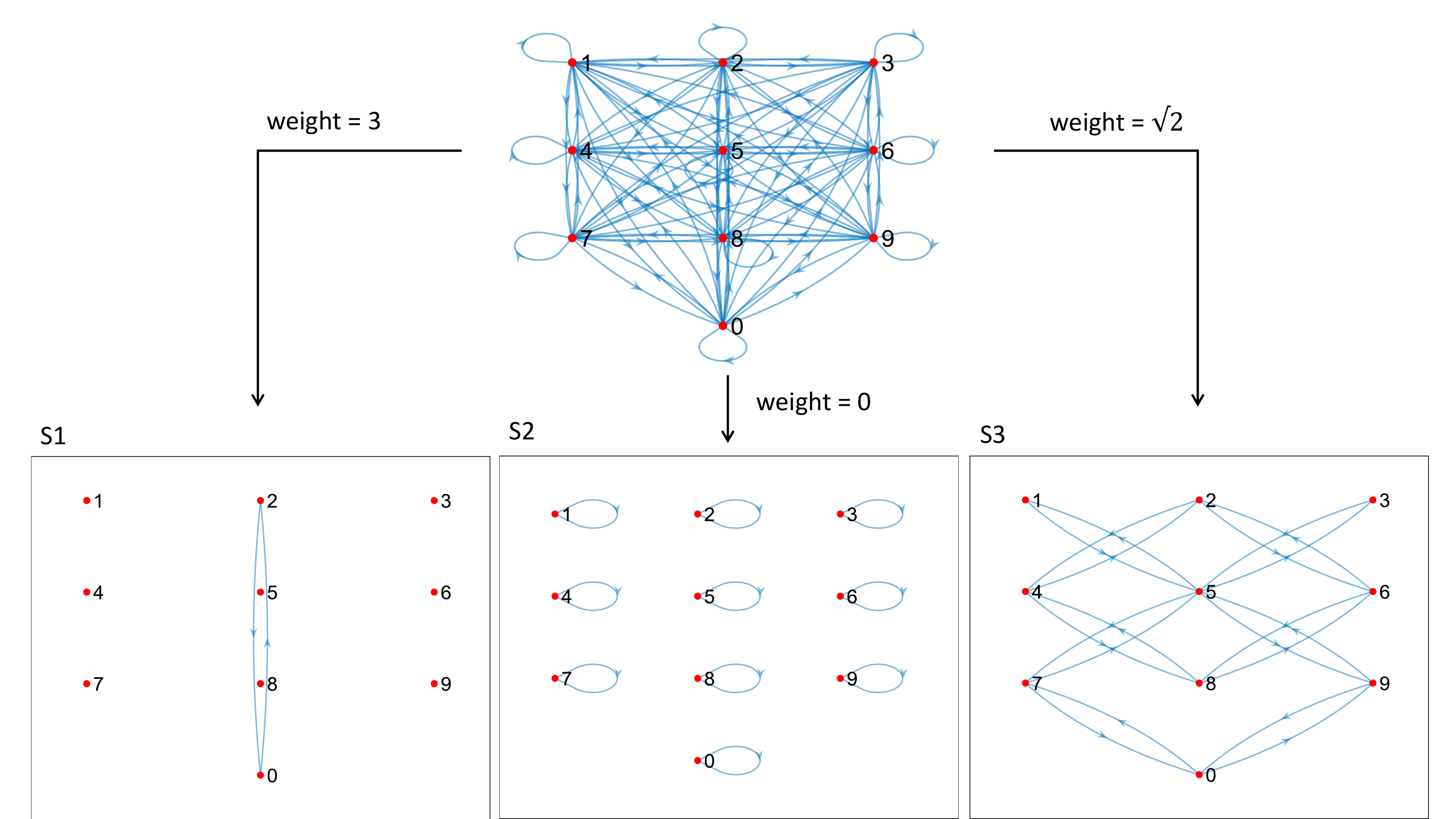}
	\caption{Full graph, and subgraphs $h_1 \in S_1$ ($weight=3$), $h_2 \in S_2$ ($weight=0$), and $h_3 \in S_3$ ($weight=\sqrt{2}$).}\vspace{-0.5cm}
	\label{fig:SUB_3_0_d1}
\end{figure}

\begin{figure}[t]
	\centering
		\includegraphics[width=0.8\linewidth]{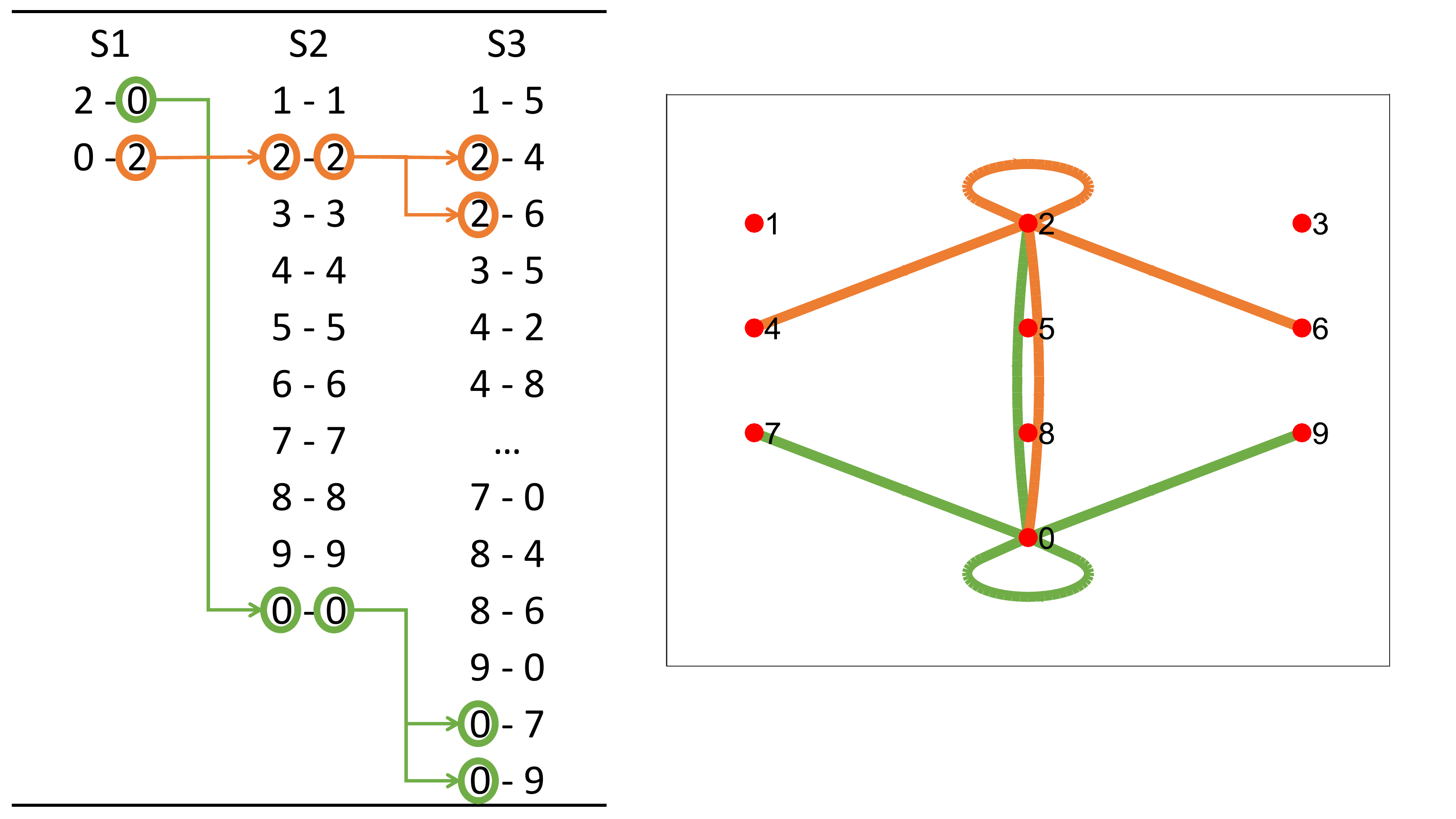}
	\caption{Nodes sequences for input distances 3, 0, $\sqrt{2}$.}\vspace{-0.5cm}
	\label{fig:NOD_3_0_d1}
\end{figure}

We combined the subgraphs in these sets by ensuring that, for $i=1$ and $i=2$, the second node of a graph from $S_i$ is the same as the first node of a graph from $S_{i+1}$. For instance, given estimated distances $3$, $0$, and $\sqrt{2}$, our algorithm extracts the subgraphs shown in Figure \ref{fig:SUB_3_0_d1}. It then refines these choices by removing all subgraphs from $S_2$ which do not have nodes $2$ and $0$ as their first node. The same rule is applied to $S_3$. The two resulting graphs are shown in Figure~\ref{fig:NOD_3_0_d1}. 

\begin{figure}[t]
	\centering
	\subfloat[][]
		{\includegraphics[width=0.8\linewidth]{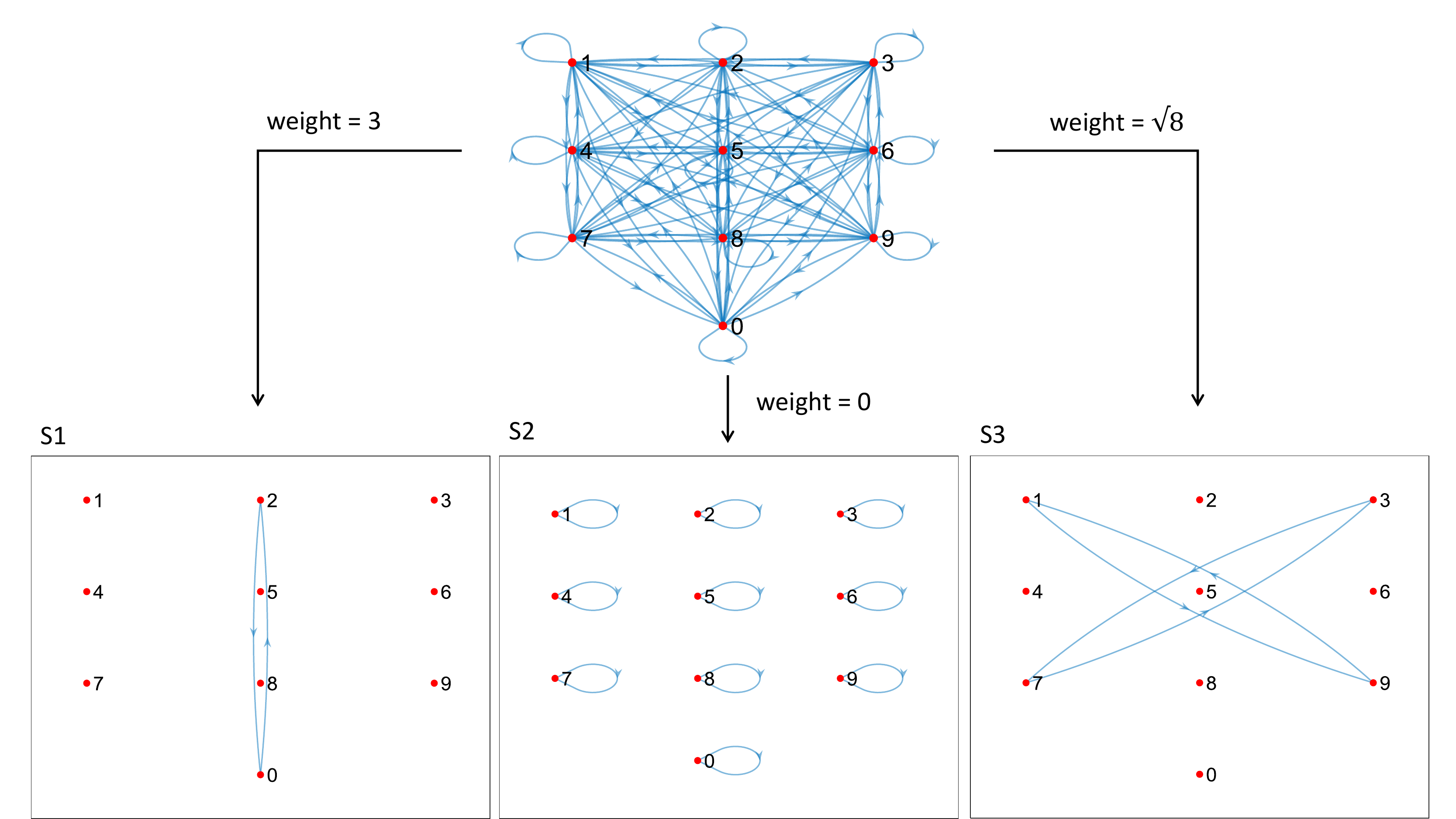}} \quad
	\subfloat[][]
		{\includegraphics[width=0.8\linewidth]{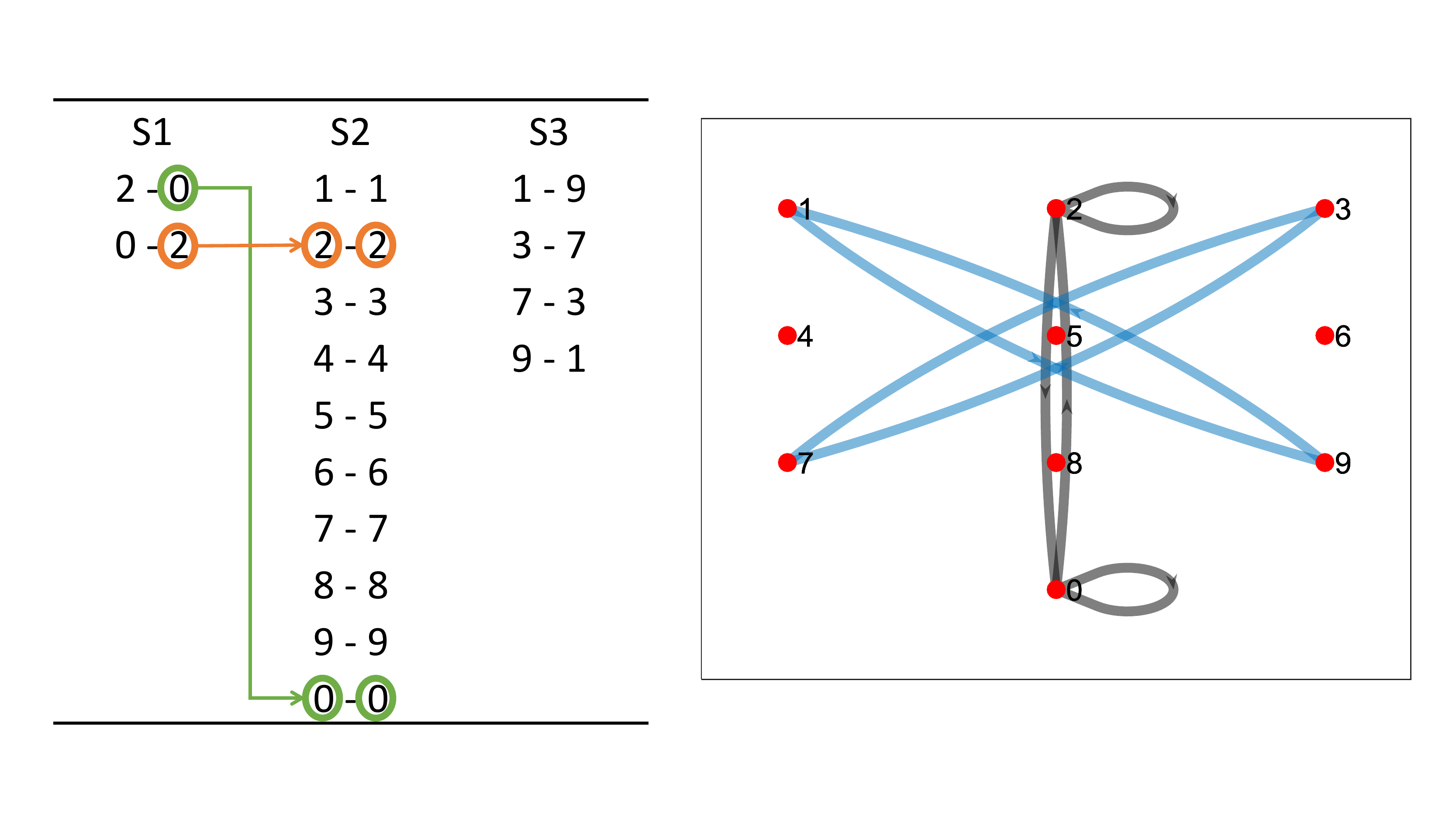}} \quad
	\caption{(a) Subgraphs corresponding to distances $3$, $0$, and $\sqrt{2^2+2^2}$. (b) Nodes connected by vertices weights 3, 0, and $\sqrt{2^2+2^2}$. No
	common nodes are in $S_2$ and $S_3$, and therefore this combination of distances does not correspond to any PIN.}\vspace{-0.5cm}
	\label{fig:SUB_NOD_3_0_d1}
\end{figure}

Not all estimated distances correspond to possible PINs. For instance, estimated distances $3$, $0$, and $\sqrt{8}$ do not match any PIN that can be typed on the pad used for the experiments: distance $3$ indicates that the second PIN digit must be either 0 or 2; as a consequence, distance $0$ associated to the second PIN digraph restricts the third PIN digit to 0 or 2; however, the set of keypairs with a relative distance of $\sqrt{8}$ (i.e., $\{(1,9), (7,3), (9,1), (3,7)\}$) does not include keys 0 or 2. Therefore, estimated distances $3$, $0$, and $\sqrt{8}$ do not lead to any valid PIN. Figure~\ref{fig:SUB_NOD_3_0_d1} shows a visual representation of this example.

For each triplet of estimated distances, the number of associated PINs may
differ. For example, 58 triplets have no associated PIN (e.g., distances $3$, $0$, and $\sqrt{8}$). The remaining
454 combinations vary from a minimum of 2 associated PINs (57 combinations;
e.g., distances $3$, $3$, and $3$ correspond only to PINs 2020 and 0202) to a maximum of 216
PINs (distances $1$, $1$, and $1$).

\begin{figure}[t]
	\centering
		\includegraphics[width=0.8\linewidth]{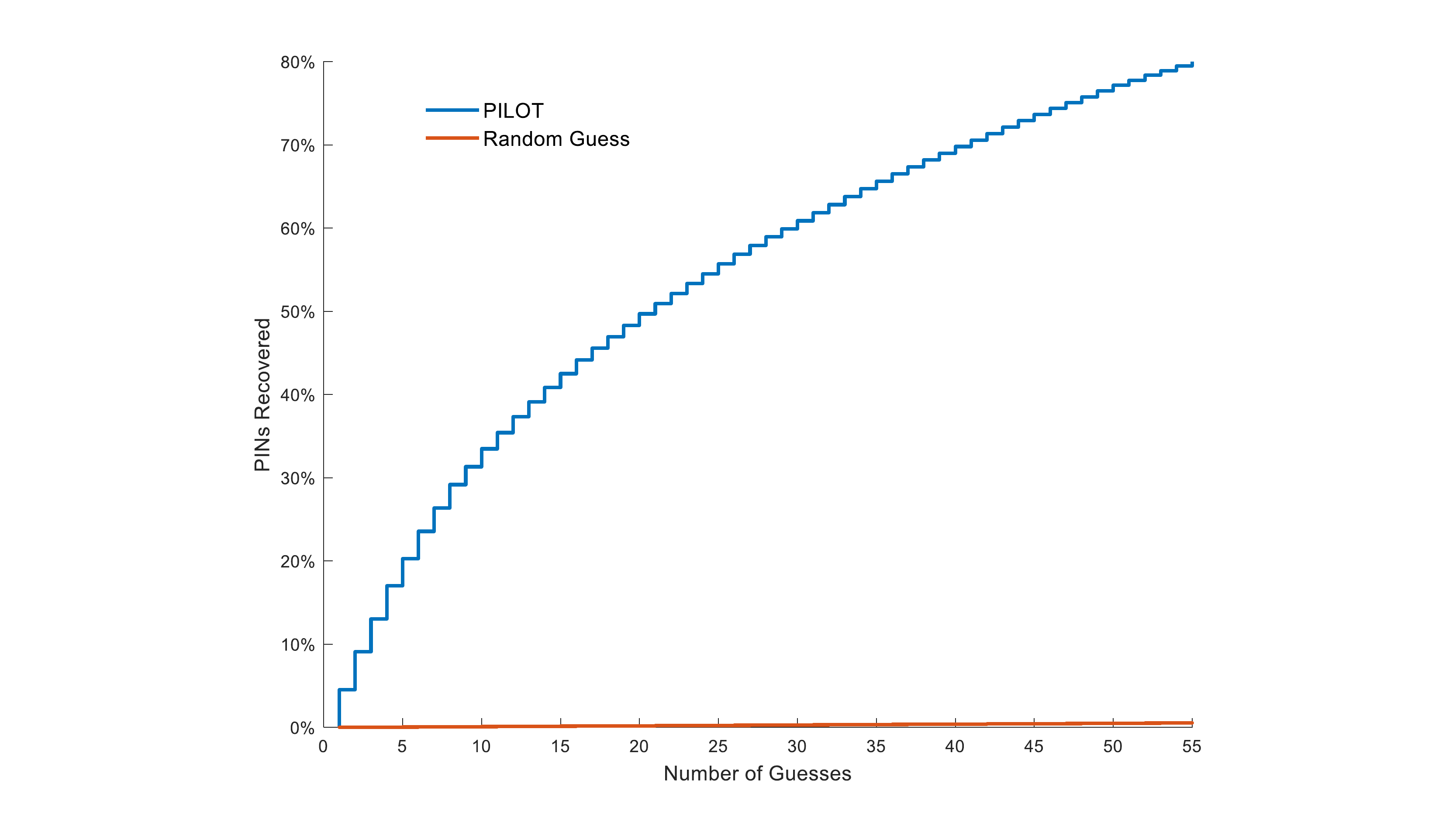}
	\caption{CDF showing the number of PINs recovered under the assumption that distances are recovered without error.}\vspace{-0.5cm}
	\label{fig:CDF_PINO_vs_Random}
\end{figure}

If the adversary is able to reconstruct the distances between digraphs without
errors, this process drastically reduces the number of attempts needed to guess
the PIN compared to a random guessing. Figure~\ref{fig:CDF_PINO_vs_Random}
shows the benefit of this approach in terms of percentage of PINs guessed within
a fixed number of attempts. However, due to the overlapping between timing
distributions shown in figures~\ref{smth} and~\ref{fig:pdf_distance1}, the
adversary cannot always estimate distances correctly. To evaluate the impact of
distance evaluation errors on PIN guessing,
we split our keystroke dataset in two sets. The
first (training set) consists of 5195 PIN, typed by 11 participants. The second (testing set) consists of 5135 PIN, typed by a distinct set of 11 participants. 

\begin{figure}[t]
	\centering
		\includegraphics[width=0.8 \linewidth]{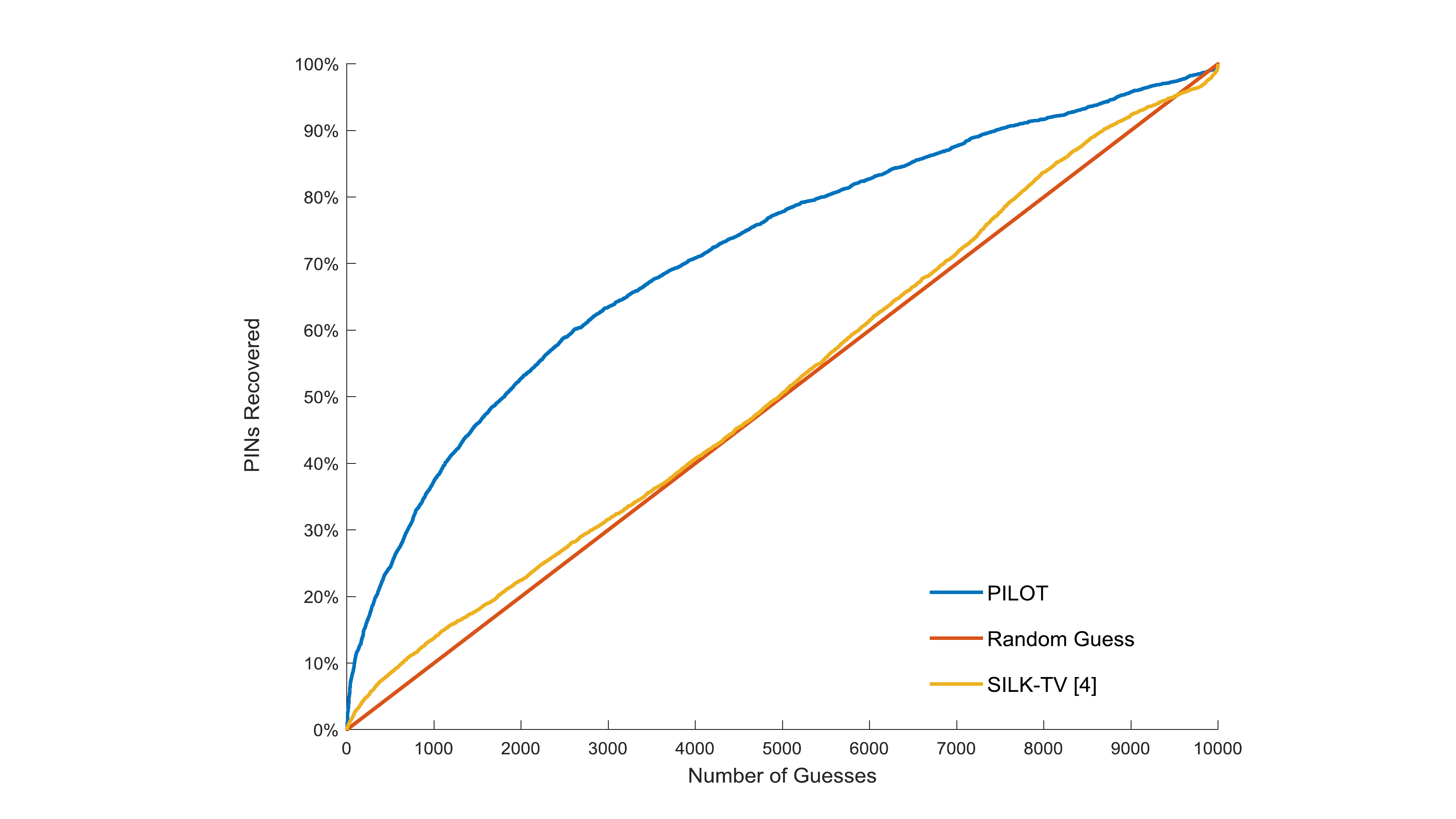}
	\caption{CDF showing the number of PINs recovered on our dataset, compared to the baseline and the algorithm of~\cite{silktv}.}\vspace{-0.5cm}
	\label{fig:Test_set_PINO_vs_Random}
\end{figure}

For each PIN in the testing set, we associated a list of triplets of distances sorted by their probability as determined using the gamma distributions.
Figure \ref{fig:Test_set_PINO_vs_Random} shows the effectiveness of this algorithm compared to random guessing, and compared to the technique used in~\cite{silktv}.
Our results show that our technique significantly improves on both random guessing and~\cite{silktv} (see Table~\ref{tab:Performance}). In particular, we were able to guess about 3\% of the PINs within 10 attempts. This corresponds to a 26-fold improvement compared to random guessing. Moreover, we were able to significantly outperform~\cite{silktv} in terms of number of PINs recovered within a small number of attempts. Specifically, we were able to recover more than 1\% of the PINs within 5 attempts, while~\cite{silktv} could guess only 0.32\% of the PINs within the same number of attempts. In summary, our technique demonstrates that inter-keystroke timings contain considerable information about the physical distance between consecutive keys in a PIN, and thus they can substantially reduce the number of attempts required to guess a PIN. 

\tabulinesep = 2mm
	\begin{table*}[h!]
		\caption{Percentage of PIN guessed using our technique, Random Guessing (RG), PIN guessing with exact distance (Exact), and \cite{silktv}}
		\label{tab:Performance}
		\begin{tabu}{lllllll}
		
			\toprule  
			& & & & & Improvement of our \\ 
			Attempts & Our work & RG & Exact & SILK-TV \cite{silktv} & work compared to RG \\

			\midrule

			 5  & 1.09\%\ & 0.05\%\ & 13.05\%\ & 0.32\%\ & 21.81$\times$\\
			 10 & 2.65\%\ & 0.10\%\ & 26.37\%\ & 0.47\%\ & 26.48$\times$\\
			 20 & 3.93\%\ & 0.20\%\ & 45.57\%\ & 0.79\%\ & 19.67$\times$\\
			 40 & 7.05\%\ & 0.40\%\ & 67.37\%\ & 1.31\%\ & 17.62$\times$\\
			 80 & 9.41\%\ & 0.80\%\ & 88.85\%\ & 2.28\%\ & 11.76$\times$\\
			 160 & 12.89\%\ & 1.60\%\ & 98.34\%\ & 3.82\%\ & 8.06$\times$\\
			 320 & 19.69\%\ & 3.20\%\ & 100.00\%\ & 6.34\%\ & 6.15$\times$\\
			 640 & 28.33\%\ & 6.40\%\ & 100.00\%\ & 10.05\%\ & 4.43$\times$\\
			 1280 & 42.30\%\ & 12.80\%\ & 100.00\%\ & 16.34\%\ & 3.30$\times$\\
			 2560 & 59.39\%\ & 25.60\%\ & 100.00\%\ & 27.81\%\ & 2.32$\times$\\
			 5120 & 78.53\%\ & 52.20\%\ & 100.00\%\ & 51.84\%\ & 1.50$\times$\\
						
			\bottomrule
		\end{tabu}
	\end{table*}

\section{Conclusion}\label{sec:conclusion}

In this paper, we have shown that inter-key timing information disclosed by
showing password masking symbols can be effectively used to reduce the cost of
password guessing attacks. To determine the impact of this side channel, we
recorded videos from 84 subjects, typing several passwords and PINs under
different conditions: in a lecture hall, while their laptop was collected to a
projector; in a classroom setting; and using a simulated ATM machine. Our
results show that: (1) it is possible to infer very accurate timing information
from videos of LCD screens and projectors (the average error was 8.7ms, which 
corresponds to the duration of a frame when the refresh rate of a display is
set to 60 Hz); (2) inter-keystroke timings reduce the number of attempts to
recover a password by 25\% and 385\%, with some passwords guessed within 19
attempts. We consider this a substantial reduction in the cost of password
guessing attacks, to the point that we believe that masking symbols should not
be publicly displayed when typing passwords; and (3) disclosing inter-keystroke
timings have a significant impact on PIN guessing attacks. In particular, we
were able to predict about 1 in 14 PINs within 40 attempts, compared to one in
250 with no timing information. While this result is not as dramatic as the one
with passwords, it suggests that keystroke timing information should be
carefully concealed by ATMs.

Clearly, the benefits of \atkname{} compared to our baseline attack vary
depending on how common the user's password is. For very common (and therefore
very easy to guess) passwords, our results show that \atkname{} might not be
needed. On the other hand, the speedup offered by \atkname{} when guessing rare
passwords is substantial. Given the effectiveness of this attack on password
guessing, we think that future work should consider countermeasures that strike
the right balance between usability and security when displaying masking
symbols. For instance, GUIs may not display masking symbols on a secondary
screen (e.g., projectors), or may display new masking symbols at fixed intervals
(say, every 250ms). Clearly, both countermeasures have usability implications,
and we leave the quantification of this impact to future work.

{\small
\subsection*{Acknowledgements}
Kiran Balagani and Paolo Gasti were supported by the National Science Foundation under Grant No. CNS-1619023. Tristan Gurtler, Charissa Miller, Kendall Molas, and Lynn Wu were supported by the National Science Foundation under Grant No. CNS-1559652. This work is partially supported by the EU TagItSmart! Project (agreement H2020-ICT30-2015-688061), and the EU-India REACH Project (agreement ICI+/2014/342-896).
}

\bibliography{bibliography}

\end{document}